\documentclass[12pt,preprint]{aastex}
\shorttitle{Specific SFR profiles of nearby disks}
\shortauthors{Mu\~{n}oz Mateos et al.}

\begin{document}

\title{Specific SFR profiles in nearby spiral galaxies: quantifying the inside-out formation of disks}

\author{J.C. Mu\~{n}oz-Mateos\altaffilmark{1},
A. Gil de Paz\altaffilmark{1},
S. Boissier\altaffilmark{2},
J. Zamorano\altaffilmark{1},
T. Jarrett\altaffilmark{3},
J. Gallego\altaffilmark{1},
B.F. Madore\altaffilmark{4}}

\altaffiltext{1}{Departamento de Astrof\'{\i}sica y CC$.$ de la Atm\'osfera, Universidad Complutense de Madrid, Avda$.$ de la Complutense, s/n, E-28040 Madrid, Spain; jcmunoz, agpaz, jaz, jgm@astrax.fis.ucm.es}
\altaffiltext{2}{Laboratoire d'Astrophysique de Marseille, BP 8, Traverse du Siphon, 13376 Marseille Cedex 12, France; samuel.boissier@oamp.fr}
\altaffiltext{3}{Spitzer Science Center, California Institute of Technology, 100-22 IPAC, Pasadena, CA 91125; jarrett@ipac.caltech.edu}
\altaffiltext{4}{Observatories of the Carnegie Institution of Washington, 813 Santa Barbara Street, Pasadena, CA 91101; barry@ociw.edu}

\begin{abstract}
We present specific Star Formation Rate (sSFR) radial profiles for a
sample of 161 relatively face-on spiral galaxies from the GALEX Atlas
of Nearby Galaxies. The sSFR profiles are derived from GALEX \& 2MASS
(FUV$-$$K$) color profiles after a proper SFR calibration of the UV
luminosity and $K$-band mass-to-light ratio are adopted. The
(FUV$-$$K$) profiles were first corrected for foreground Galactic extinction 
and later for internal
extinction using the ratio of the total-infrared (TIR) to FUV
emission. For those objects where TIR-to-FUV-ratio radial profiles
were not available, the (FUV$-$NUV) color profiles as a measure of the
UV slope. The sSFR radial gradients derived from these profiles allow
us to quantify the inside-out scenario for the growth of spiral disks
for the first time in the local Universe.

We find a large dispersion in the slope of the sSFR profiles with a
slightly positive mean value, which implies a moderate inside-out disk
formation. There is also a strong dependency of the value of this
slope on the luminosity and size of the disks, with large systems
showing a uniform, slightly positive slope in almost all cases and
low-luminosity small disks showing a large dispersion with both
positive and negative large values. While a majority of the galaxies
can be interpreted as forming stars gradually either from inside out
or from outside in, a few disks require episodes of enhanced recent
growth with scale lengths of the SFR (or gas infall) being
significantly larger at present than in the past. We do not find any
clear dependence of the sSFR gradient on the environment (local galaxy
density or presence of close neighbors).

\end{abstract}


\section{Introduction}

According to the $\Lambda$CDM paradigm of hierarchical galaxy
formation the inner parts of galactic disks form first followed by the
formation of their outer regions (White \& Frenk 1991; Mo, Mao, \&
White 1998). This naturally results in a gradual growth of the size
(i.e$.$ scale-length) of the disks with time. This {\it inside-out}
formation scenario has been also confirmed by recent N-body/SPH
simulations of the evolution of individual disk galaxies (e.g$.$ Brook
et al. 2006).

Moreover, the inside-out scenario of disk formation has been also
proposed to explain the radial variation of the abundances of elements
and colors in the disk of our own Milky Way (Matteucci \& Francois
1989; Boissier \& Prantzos 1999). In the case of the models for our
Galaxy, the inside-out scenario is usually taken into account by
increasing the gas-infall timescale with radius. The results of such
models are in agreement with observables in our Milky Way such as the
abundance gradients, and the wavelength dependence of the scale-length
of the disk. Interestingly, this radial increase in the gas-infall
timescale has been reproduced by some recent N-body/SPH simulations by
Sommer-Larsen et al. (2003; model S1) and Samland \& Gerhard
(2003). Another independent result that is consistent with this
scenario comes from the weak dependence found of the mass-size
relation of distant disk galaxies with redshift, since according to
the inside-out growth of disks the scale-length is expected to
increase roughly proportional to the stellar mass (Barden et al$.$
2005; Trujillo et al$.$ 2004, 2006).

In spite of this being a long-known prediction of both hierarchical
and chemical evolution models of galaxy formation very few
observational data have been brought forward to convincingly support
it. This is probably consequence of the fact that determining the
sizes of disks of intermediate redshift has traditionally been a
complicated task due to the cosmological surface brightness dimming,
band shifting, and to the problems for identifying the distant
counterparts to the population of local disk galaxies. Recently,
Trujillo \& Pohlen (2005) have proposed using the truncation radius of
galaxies at different redshifts as a measure of the growth rate in
galactic disks. Using the data from the Hubble Ultra Deep Field, these
authors have estimated that galactic disks have suffered a
small-to-moderate growth of $\sim$25\% since redshift $z$=1.

In the same way that it has been done in the Milky Way, the analysis
of the color profiles of nearby spiral galaxies might provide
important clues to determine if the bulk of the galatic disks have
indeed formed from inside out. Radial variations of the SFH have been
identified as a key mechanism to explain color gradients in disks (de
Jong 1996). These gradients can be interpreted on the basis of
different scale-lenghts of the disk in different bands, a result which
is predicted by models based on the inside-out scenario (Prantzos \&
Boissier 2000). The study of color gradients in nearby spirals has
been addressed by several authors (see e.g$.$ de Jong 1996; Taylor et
al$.$ 2005), mainly in the optical range. In order to better relate
color gradients to a radially varying SFH, observations at other
wavelengths are needed. In particular, the comparison between the
ultraviolet-light profiles, very sensitive to the presence of recent
star formation activity, and those in the near-infrared $K$-band,
sensitive to the accumulated star formation in the galaxy, would give
a direct measure of the recent disk growth. In this paper we derive
the azimuthally-averaged radial profiles in (FUV$-$$K$) color (or,
equivalently, specific Star Formation Rate, sSFR hereafter; see
Appendix~A) for a sample of 161 relatively face-on nearby
($d<200$\,Mpc) spiral galaxies as a metric tracer of the disk
growth. Throughout this article we have adopted a concordant
cosmology, with $H_{0}=70$ km/s/Mpc, $\Omega_{M}=0.3$ and
$\Omega_{\Lambda}=0.7$.

In Section~2 we present the sample of galaxies for which the
availability of both ultraviolet and near-infrared allow us to
determine sSFR profiles. The analysis methods are described in
Section~3. Results are presented in Section~4. Assuming a simple
formulation for the star formation history of the disks of these
galaxies in Section~5 we introduce a model that relates the slope of
the sSFR profiles to the growth of the disks. A comparison between the
predictions of this model and the sSFR profiles measured is discussed
in Section~6. Finally, we summarize our conclusions in Section~7.

\section{The sample}

We have compiled a sample of 161 galaxies from the GALEX Ultraviolet
Atlas of Nearby Galaxies (Gil de Paz et al$.$ 2006) on the basis of
two main selection criteria: morphological type and inclination. We
limit our sample to spiral galaxies with Hubble types from S0/a to Sm,
i.e$.$ galaxies with $-0.5 \leq T \leq 9.5$ according to the Third
Reference Catalogue of Bright Galaxies (RC3; de Vaucouleurs et al$.$
1991).

In order to minimize the effects of internal extinction and ambiguity
in the morphological class, we select only moderately face-on galaxies
($i \leq 45^{\circ}$). We compute the inclination angle using the
major and minor axis sizes at the 25$^{\mathrm{th}}$ magnitude
isophote in the B band, as given in the RC3. The intrinsic thickness
of the disk can be taken into account by using the expression
$\cos^{2} i = (q^{2}-q^{2}_{0})/(1-q^{2}_{0})$, where $q$ is the
observed minor to major axis ratio and $q_{0}$ the intrinsic
flattening of the disk as seen edge-on. The latter has been estimated
as a function of morphological type by Guthrie (1992). Since no value
of $q_{0}$ is given for type Sdm in this paper, we assume a
intermediate value of 0.25.

These two general constraints yield an initial subset of 178
galaxies. It should be noted, however, that the inclination angle
derived from the semiaxis ratio may not be reliable for galaxies with
asymmetric arms, due to the interaction with a companion galaxy
(e.g$.$ M51a), or for those ones with S-shaped arms coming out of a
central ring or bar (e.g$.$ M95, NGC1097). In these cases the
inclination angle computed from the semiaxis ratio is usually
overestimated. Consequently, we visually inspected the whole GALEX
Atlas looking for galaxies that, in spite of being clearly face-on, do
not match our initial selection criteria. We also added to our sample
a few extensively-studied galaxies whose inclination angles are
slightly above our $45^{\circ}$ limit (M33, NGC0300). In summary, a
total of 23 additional galaxies were included in our sample.

Due to the fact that GALEX FUV detector had to be turned off during
periods of unusual solar activity or overcurrent events to preserve
the detector electronics, some galaxies in the Atlas were observed
only in the NUV band. Thus, we had to remove from our sample 26
galaxies that lacked FUV images. Another 14 objects do not have
available 2MASS data and were also excluded from the final sample,
which is constituted by 161 disk galaxies.

In Table~\ref{table1} we present basic data of the galaxies in our
sample. The equatorial coordinates were taken from NED, and typically
are derived from the 2MASS position. We show the apparent major
diameter of the elliptical isophote measured at $\mu_{B}$ = 25 mag
arcsec$^{-2}$, D25, as well as the morphological type, both taken from
the RC3 catalog. The Galactic color excess is derived from the maps of
Schlegel et al$.$ (1998). The inclination angle is obtained as
explained above. Apparent magnitudes in the FUV band are asymptotic,
i.e$.$ they were computed by extrapolating the growth curve using the
surface brightness profiles in FUV (Gil de Paz et al$.$ 2006). For the
$K$ band, apparent total magnitudes ($K_{\mathrm{tot}}$) from the
2MASS Large Galaxy Atlas (LGA, Jarrett et al$.$ 2003) were used when
available, and those given in the 2MASS Extended Source Catalog (XSC;
Jarrett et al. 2000) were adopted for those galaxies not in the
LGA. Distances to each object were compiled from a wide variety of
resources (see Gil de Paz et al$.$ 2006 for details).

\section{Analysis}\label{analysis}

\subsection{Radial profiles}\label{radialprofs}

Surface brightness radial profiles at both GALEX UV bands have been
presented in Gil de Paz et al$.$ (2006). They were obtained with the
IRAF\footnote{IRAF is distributed by the National Optical Astronomy
Observatories, which are operated by the Association of Universities
for Research in Astronomy, Inc., under cooperative agreement with the
National Science Foundation.} task \texttt{ellipse} by measuring the
mean intensity and \textit{rms} along elliptical isophotes with fixed
ellipticity and position angle, equal to those of the $\mu_{B}$ = 25
mag arcsec$^{-2}$ isophote from the RC3 catalog. The center of these
ellipses were set at the coordinates shown in Table~\ref{table1}, with
a constant increment of 6 arcsec along the semimajor axis to a final
radius at least 1.5 times the D25 radius. In the final profiles, the
outermost points were removed when the intensity fell below the
level of the sky, or when the error in the surface photometry in the
NUV band was larger than 0.8\,mag.

In order to derive color profiles in a consistent way we used the same
set of elliptical isophotes to obtain surface brightness profiles in
the $K$ band. Near-infrared images for 121 galaxies (75\% of our
sample) were compiled from the 2MASS XSC. The remaining 40 objects are
too large to fit into a single 2MASS scan or lay too close to an
edge. In those cases individual mosaics were obtained from the 2MASS
LGA. All FITS were already background and star substracted; however,
in some of the XSC images the star-substraction algorithm failed to
detect some stars, which had to be masked by hand. Companion galaxies
were also masked before measuring the profiles. As for the GALEX
images, foreground stars were detected and masked as those point
sources having $(FUV-NUV)$ colors redder than 1\,mag; these masks were
later modified after a visual inspection. Detailed information about
this process can be found in Gil de Paz et al$.$ (2006).

Due to the fact that for most galaxies, as we show in
Section~\ref{sSFRprof}, surface brightness in the $K$ band decreases
faster than in FUV as we move away from the center of the galaxy and
because the 2MASS images are not very deep ($K<20$\,mag
arcsec$^{-2}$), our $K$-band surface brightness profiles are usually
restricted to a somewhat smaller radius than those in the FUV. For
M33, we made use of a deeper (6X) 2MASS image ($K<21$\,mag
arcsec$^{-2}$; see also Block et al$.$ 2004). M33 is close enough to
resolve individual stars, but discerning them from foreground stars in
the Milky Way is not straightforward. In fact, the star removal was
performed in a statistical way, comparing $(J-K)$ color histograms of
adjacent control fields and M33 itself (see Block et al$.$ 2004 for
details). Therefore, although the star removal is correct from a
statistical point of view, some stars in M33 might have been removed
and vice versa, some sources still visible in the final image could be
foreground stars. After examining its (FUV$-$$K$) profile, we
concluded that data points more than 25\,arcmin from the galaxy center
(measured along the semimajor axis) were highly contaminated by
foreground stars, with little or no contribution coming from M33
sources, and were drawn out from the profile.

Uncertainties in the surface photometry were derived following the
prescriptions given by Gil de Paz \& Madore (2005), as had been
already done for the FUV profiles (Gil de Paz et al$.$ 2006). The
random error in $\mu_{K}$ depends on the intrinsic variation of the
intensity within each elliptical isophote and the error in the sky
level. The latter includes two different contributions: Poisson noise
in the sky level (as well as pixel-to-pixel flat-fielding errors), and
low spatial frequency flat-fielding errors. Considering that in 2MASS
images flat-field correction and sky-substraction are carried out
using whole scans, we can safely assume that the contribution of
low-frequency errors to the final uncertainty is negligible compared
to that of the high-frequency ones. Hence, we compute the final random
uncertainty in $\mu_{K}$ from the \textit{rms} along each isophote
(which is part of the output of the IRAF task \texttt{ellipse}) and
local noise measurements provided as part of the headers of the
XSC/LGA FITS files.

Once the mean surface brightness and its error have been measured,
they are converted into magnitudes/arcsec$^{2}$ using the calibrated
zero point provided in the header of each XSC/LGA FITS image. This
introduces a systematic uncertainty of $\pm$0.007\,mag in the $K$-band
magnitudes, which is considerably smaller than the zero-point error of
0.15\,mag in the FUV data.

Before combining our $K$-band profiles with those in FUV, we correct
them for foreground Galactic extinction using the color excesses from
the Schlegel et al$.$ (1998) maps and the parameterization of the
Galactic extinction curve given by Cardelli et al$.$ (1989). We assume
a value of $R_{V}=3.1$, which implies a total-to-selective extinction
in the $K$ band of $A_{K}=0.367 \times E(B-V)$ [compare with
$A_{FUV}=7.9 \times E(B-V)$].

Fig.~1 shows the (FUV$-$$K$) color profiles for six typical galaxies
in our sample (upper region of each panel). Gray points are only
corrected for Galactic extinction, as described above, whereas black
points are also corrected for internal extinction (see Section
\ref{extinction}). Each point is shown with its corresponding error,
computed as the square-root of the quadratic sum of random
uncertainties in both $\mu_{K}$ and $\mu_{FUV}$. For the sake of
clarity, error bars do not include the systematic zero-point
uncertainty in (FUV$-$K), which is $\sim$0.15\,mag (the zero-point
error in FUV dominates over the one in $K$). Note that this zero-point
uncertainty does not affect to the shape of the profiles, but only the
normalization. Moreover, since the vast majority of the images in the
GALEX Atlas were processed and reduced following the same version of
the GALEX pipeline, the effect of the zero-point error is expected to
be the same for all our galaxies. The lower plot in each panel shows
the (FUV$-$NUV) color profile for each galaxy (see
Section~\ref{extinction}).

\subsection{Internal extinction correction}\label{extinction}

Before computing the specific SFR as a function of radius, the
(FUV$-$$K$) color profiles must be corrected for internal
extinction. The ultraviolet light emitted by young massive stars is
absorbed and scattered by dust, and then remitted in the far-infrared
(FIR). Hence, the ratio of far-infrared to ultraviolet luminosity is
directly related to dust extinction. Furthermore, on the basis of the
results from previous works (e.g$.$ Buat et al$.$ 2005 and references
therein), this ratio has proven to be only weakly dependent on certain
intrinsic properties of galaxies, such as the internal extincion law,
the spatial distribution of dust and stars, or the galaxy star
formation history. Therefore, although FUV radiation can be highly
attenuated by dust, it is possible to infer $A_{FUV}$ and hence
recover the emitted FUV luminosity, and from that the SFR (Kennicutt
1998).

We correct our color profiles using the prescriptions in Boissier et
al$.$ (2006). Using IRAS data for a sample of well resolved objects in
the Galex Atlas of Nearby Galaxies, Boissier et al$.$ derived FIR and
total-infrared (TIR) profiles from measurements at 60\,$\mu$m and
100\,$\mu$m. Once combined with the UV profiles, the radial profiles
in the $L_{TIR}/L_{FUV}$ luminosity ratio (or TIR-to-FUV ratio)
obtained are converted into $A_{FUV}$ profiles using the polynomial
fits of Buat et al$.$ (2005).

We have applied this internal extinction correction to 16 galaxies in
our sample that were also studied by Boissier et al$.$ (2006). Since
the images employed to derive extinction profiles were degraded to
match the IRAS resolution, we had to interpolate $A_{FUV}$ values for
our radial profiles. In order to achieve a smooth result and avoid
artifacts in our corrected color profiles, a spline interpolation
method was used for all galaxies except for NGC1291, where the
extinction curve was not smoothly reproduced by the interpolation
algorithm and a linear interpolation was adopted instead. The upper
and lower uncertainties in $A_{FUV}$ usually change more abruptly with
radius than $A_{FUV}$ itself, thus making the spline interpolation
unreliable. Therefore, these errors were linearly interpolated and
then assigned to their corresponding $A_{FUV}$ values.

However, for the majority of galaxies in our sample we cannot apply
the direct extinction correction based on the $L_{TIR}/L_{FUV}$
profiles. Fortunately, the infrared excess $L_{TIR}/L_{FUV}$ (and
hence $A_{FUV}$) is related to the slope of the UV spectrum or,
equivalently, the (FUV$-$NUV) color, with redder (FUV$-$NUV) colors
meaning higher dust attenuation (see e.g$.$ Cortese et al$.$ 2006 and
references therein). Although this relation (known as the IRX-$\beta$
law) was originally found to be applicable only to actively
star-forming systems (Meurer et al$.$ 1999), recent work by Gil de Paz
et al$.$ (2006) show that, although with a significant dispersion,
such a trend is also present in normal spiral galaxies such as those
in our sample, especially when the TIR-to-FUV and (FUV$-$NUV) radial
profiles of the disks of these galaxies are compared (Boissier et
al$.$ 2006). Here we have taken advantage of the empirical relation
between the TIR-to-FUV ratio and the (FUV$-$NUV) color derived by
Boissier et al$.$ (2006) and the (FUV$-$NUV) radial profiles presented
by Gil de Paz et al$.$ (2006). This procedure allows us to obtain
$A_{FUV}$ profiles for the remaining 145 galaxies in our sample. The
uncertainties in $A_{FUV}$ were computed from the upper and lower
limits of the 1$\sigma$ prediction band for Boissier et al$.$ fit to
the IRX-$\beta$ plot.

Although to a significantly reduced extent, the $K$-band luminosity is
also affected by the presence of dust inside the disk of the
galaxy. Assuming the Cardelli et al$.$ parameterization of the
extinction curve to be valid in our galaxies, we compute the K-band
extinction to be $A_{K}=0.0465 \times A_{FUV}$. The choice of a
different internal extinction curve would have not significantly
affected our extinction corrected $\mu_{K}$ profiles since $A_{K}\ll
A_{FUV}$ independent of the composition and physical properties of
dust grains.

Table~\ref{table2} shows the color profiles for the galaxies in our
sample. Since we are dealing with disk-like galaxies, we use the
radius along the semi-major axis of the elliptical isophotes instead
of the equivalent radius (as it was done in Gil de Paz et al$.$ 2006)
to describe our profiles. We show the measured FUV and $K$-band
surface brightness as well as the (FUV$-$NUV) color, corrected only
for foreground Galactic extincion as described in Section
\ref{radialprofs}. The different values of $A_{FUV}$ and the corrected
(FUV$-$$K$) profiles are also given, with their corresponding upper
and lower uncertainties. Finally, we give the specific SFR at each
radius for each galaxy (see Section \ref{sSFRprof} for the derivation
of the sSFR).

Fully corrected profiles can be seen in the upper part of each panel
in Fig.~1 as black dots, whose error bars account for both photometric
and extinction-correction uncertainties. The bottom part of each panel
shows the (FUV$-$NUV) color profiles.

Since the extinction correction may be subject of large uncertainties
(both random and systematic), it is important to determine to what
extent the observed (FUV$-K$) color profiles are determined by radial
variations in the extinction rather than by the intrinsic colors of
the underlying stellar population (and hence by the SFH). In order to
address this important issue, in Fig.~2 we plot $m_{A_{FUV}}$, the
radial gradient of the extinction in the FUV (see also
Table~\ref{table3}), against $m_{(FUV-K)\mathrm{obs}}$, the
\textit{observed} (FUV$-K$) color gradient, both of them measured in
the disk-dominated region of the profiles (that is, excluding the
bulge). The cross shows the mean uncertainties in both parameters
($\Delta m_{(FUV-K)\mathrm{obs}} \sim 0.06$\,mag/kpc and $\Delta
m_{A_{FUV}} \sim 0.05$\,mag/kpc). Note that almost 30\% of the sample
show errors lower than half these values, for which the derived sSFR
profiles will be most reliable. See Section~\ref{results} for an
in-depth description of the fitting procedure.

The diagram has been divided into four zones. The diagonal line is the
loci of galaxies with $m_{A_{FUV}}=m_{(FUV-K)\mathrm{obs}}$, meaning
that the intrinsic (FUV$-K$) profile is flat and the observed color
gradient is entirely due to radial changes in the
extinction. Therefore, galaxies to the left of this line can be
described in terms of an inside-out formation, with the stellar
population becoming relatively bluer and younger with increasing
radiues, and vice versa. On the other hand, galaxies in the lower half
of the figure are those in which the dust content decreases with
radius, while those in the upper half have positive dust gradients
(note that the limitations of the IRX-$\beta$ plot may constitute an
important caveat here).

Two important conclusions concerning the extinction correction can be
derived from this plot. Most galaxies are located in the bottom-left
region of the plot, as would be expected, and they do not follow the
diagonal line, meaning that we can actually obtain reliable sSFR
gradients since the observed color gradient is not only due to
variations of A$_{FUV}$. Secondly, correcting for internal extinction
is clearly essential to properly study the evolution and growth of
disks from color gradients. Had we simply used the observed (FUV$-K$)
color profile to compute the specific SFR, the boundary between
inside-out and outside-in scenarios would have been a vertical line at
$m_{(FUV-K)obs}=0$ rather than a diagonal one, leading to an
overestimation of their inside-out growth.

\section{Results}\label{results}

\subsection{Specific SFR profiles}\label{sSFRprof}

Once the Galactic and internal extinction corrections have been
applied, we proceed to compute the specific Star Formation Rate
(sSFR). Following the calibration given by Kennicutt (1998) to convert
FUV luminosity into SFR, the specific SFR can be expressed as a
function of (FUV$-$$K$) as
\begin{equation}
\log(sSFR)(\mathrm{yr}^{-1})=-0.4(FUV-K)-8.548-\log(M/L_{K})\label{eqsSFR}
\end{equation}
where $M/L_{K}$ is the stellar mass-to-light ratio (expressed in solar
units) in the K-band (see Appendix A). We have adopted a constant
value of $M/L_{K}=0.8 M_{\sun}/L_{\sun ,K}$ (Bell et al$.$ 2003)
across the entire extent of the disk. Indeed, the choice of a
different mass-to-light ratio would not modify the radial gradient of
the specific SFR, only the global normalization, as long as it remains
constant all over the disk. However, the mass-to-light ratio could
depend on the galactocentric distance; the effects that radial
variations of $M/L_{K}$ could have on the sSFR gradient are discussed
later in this section.

It is widely known that light profiles of disks usually follow an
approximately exponential law. Therefore, (FUV$-$$K$) and sSFR (once
expressed in log scale) are also expected to track linearly with
radius. Thus, in order to characterize the main properties of the sSFR
radial variations we have performed a linear fit to our sSFR profiles
(in log scale). Obviously, local features in the profiles due to rings
or arms etc., cannot be described with a straight line; we are mainly
interested in the global gradient of sSFR along the whole disk, over
spatial scales somewhat larger than these local features.

We must first exclude the bulges from this analysis for several
reasons. The extinction correction we have applied to our profiles is
only valid in star-forming systems like the disks of spiral
galaxies. Bulges and early-type galaxies do not usually host active
star formation processes, and their red (FUV$-$NUV) colors are mainly
due to the intrinsically red underlying stellar population rather than
dust attenuation. Besides, evolved giant stars in the blue part of the
horizontal branch may contribute to the observed UV flux. Finally, in
galaxies with AGN activity a blue peak could be observed associated to
the innermost nuclear regions.

The radius at which the contribution of the bulge to the (FUV$-$$K$)
color is negligible compared to that of the disk ($r_{in}$ hereafter)
was determined by visually inspecting the (FUV$-$NUV) radial profiles,
complemented with UV and optical images. In most cases the bulge-disk
separation is rather easy to determine, since both (FUV$-$K) color and
its gradient experience an obvious change. On the other hand, we found
many galaxies with no apparent bulge or a very small one. For some of
these objects the size of the bulge is of the order of the resolution
of the GALEX images (PSF FHWM$\simeq$5\,arcsec). For those galaxies we
adopted a conservative criterion ($r_{in}$=9\,arcsec) and removed the
first point of the profiles from the linear fit, just to be sure that
the contribution of the bulge or AGN (if present) does not
significantly alter our results.

Table~\ref{table3} shows the value of $r_{in}$ used for each galaxy,
as well as the resulting parameters of the linear regression: the
extrapolated value of sSFR at $r=0$ ($sSFR_{0}$ hereafter) and the
slope ($\Delta \log(sSFR)/\Delta r$, $m_{sSFR}$ hereafter). These
values are obtained by performing an unweighed linear fit over the
central (i.e$.$ most probable) values of log(sSFR) (column 7 in
Table~\ref{table2}). Traditionally, weighted linear fits only take
into account the relative weight of each point with respect to the
others based on their errors but not the absolute value of each
individual uncertainty. Although this results in a correct estimate of
the best-fitting parameters and their variances when the individual
uncertainties are comparable to the dispersion around the best-fit, it
might lead to wrong estimates if either individual uncertainties are
much smaller than the dispersion of the data (due to the presence of
outliers) or the opposite, individual errors are much larger than the
actual dispersion of the data (Press 1992). This is actually the case
for those of our sSFR profiles corrected for extinction using the
empirical relationship between (FUV$-$NUV) color and TIR-to-FUV ratio
given by Boissier et al$.$ (1996). The use of this relationship
results in relatively large uncertainties in the corrected (FUV$-$$K$)
color profiles but small dispersion between individual data
points. This is a consequence of assuming that the value of the
TIR-to-FUV ratio of the individual data points in the profile for a
given galaxy can be found anywhere within the 1-$\sigma$ prediction
band of Boissier et al$.$ (2006) for a given (FUV$-$NUV) color, i.e$.$
the internal-extinction corrections are independent from point to
point. This behavior in the data results in errors in the parameters
that are clearly underestimated if a standard weighted least-squares
fitting technique is used.

Consequently, in order to derive more realistic errors for $m_{sSFR}$
and $sSFR_{0}$ we performed Monte Carlo simulations on our
profiles. For each data point in a profile we generated 2000 random
points following a normal distribution with $\mu=0$ and $\sigma=1$,
which are then rescaled according to the upper and lower uncertainties
of log(sSFR) and added to the central values (assumed to be the most
probable ones). We then apply a linear fit to each new random profile,
ending up with a set of 2000 values of $sSFR_{0}$ and $m_{sSFR}$. We
finally compute the upper and lower standard deviations of these
randomly obtained values with respect to our best-fit values
previously derived. These uncertainties are shown in Table 3.

In order to compute the specific SFR from the (FUV$-$$K$) color we
make the assumption that the $K$-band mass-to-light ratio depends only
weakly on the galactocentric distance and a constant
$M/L_{K}=0.8M_{\sun}/L_{\sun ,K}$ can be adopted. As discussed by Bell
\& de Jong (2001), for a stellar population showing a wide range of
optical colors [$(B-R)$=0.8 to 1.4\,mag] and timescales of formation
(from 3\,Gyr to $\infty$) the $\log(M/L_{K})$ is found to vary by only
$\sim$0.2 dex. In those disks where the $M/L_{K}$ would decrease
towards the outer and, consequently, bluer parts of the galaxy the
gradient of specific SFR would obviously be positive and slightly
larger than that derived assuming a constant $K$-band mass-to-light
ratio. A simple estimate shows that $m_{sSFR}$ could be about 0.02
dex/kpc higher in these disks for a typical radius of 10 kpc, and
possibly less for bigger galaxies, under the conservative assumption
that the timescale of formation changes from 3\,Gyr to $\infty$ across
the disk. The opposite would occur in galaxies with an opposite color
gradient, whose sSFR slopes could be reduced by a similar amount. It
is worth noting that this systematic uncertainty is significantly
smaller than our typical errors in $m_{sSFR}$ ($\sim 0.03$\,dex/kpc),
which include both photometric and extincion-correction
uncertainties.

The $M/L_{K}$ ratio can also vary with Hubble type, although this
would only affect $log(sSFR_{0})$, but not $m_{sSFR}$. Portinari,
Sommer-Larsen \& Tantalo (2004) estimate that $M/L_{K}$ can change
from around 1 to 0.6 from early to late-type spiral galaxies. These
variations around our adopted average value of $M/L_{K}=0.8$ would
globally increase the specific SFR of late-type spirals by $\sim
0.2$\,dex, and decrease it by a similar amount for early-type disks.

The conversion factor between FUV luminosity and SFR given by
Kennicutt (1998) is computed assuming solar metallicity (see Madau,
Pozzetti \& Dickinson, 1998). For a given SFR the FUV luminosity is
expected to decrease with increasing metallicity, due to the
blanketing effect caused by metallic absorption lines in the
FUV. Hence, radial metallicity gradients in our disks might constitute
an additional source of systematic uncertainty. We made use of the
Starburst99 synthesis code (Leitherer et al$.$ 1999) to estimate how
the FUV luminosity would change with different metallicities (ranging
from 0.008 to 0.040) and a fixed SFR. The FUV luminosity is found to
be $\sim0.2$\,mag fainter for $Z=0.040$ than for $Z=0.008$. Therefore,
according to Eq.~\ref{eqsSFR}, the specific SFR would be $\sim
0.08$\,dex \textit{higher} for a high-metallicity region than for a
low-metallicity one with the same FUV luminosity. Since metallicity is
usually found to decrease with the galactocentric radius, if we assume
that these extreme metallicity gradients are spread across the whole
disk ($\sim10$\,kpc), our sSFR gradients derived assuming a constant
solar metallicity within the whole disk would be overestimated by
0.01\,dex/kpc, which is still below our typical quoted
uncertainties. To summarize, we estimate that the possible spread in
$M/L_{K}$ and metallicity is not a significant component of the total
sSFR uncertainty.

The SFR calibration could be also affected by FUV radiation coming
from stars in the horizontal branch (HB), specially in the innermost
regions of our profiles. HB stars are thought to constitute a major
source of FUV radiation in elliptical and lenticular galaxies. In Gil
de Paz et al$.$ (2006) it was shown that these early-type galaxies are
usually redder than FUV$-K\simeq 9$, while spiral and irregular
galaxies typically exhibit bluer FUV$-K$ colors, since their FUV
luminosity is dominated by star formation. By comparing the FUV$-K$
colors of the innermost points of our disk profiles (the ones
immediately after $r_{in}$) with those of E and S0 galaxies presented
in Gil de Paz et al$.$ (2006), we conclude that only 21 galaxies (13\%
of our total sample) present innermost regions red enough to overlap
with the colors of elliptical and lenticular galaxies, a fraction that
decreases to nearly zero when we apply internal-extinction
corrections. Even if some contamination from HB stars might be found
in the innermost zones of those disks, the global fits to the whole
disks should not be affected by some SFR calibration changes in those
points.

\subsection{Stellar mass surface density profiles}\label{massprof}

Given the relation between stellar mass and K-band luminosity, another
interesting parameter than can be derived from the $K$-band surface
brightness profiles is the surface mass density scale-length of the
disk (i.e$.$ the radius at which the stellar mass surface density decays by a
factor $e$ with respect its value at the center; $\alpha_{M}$
hereafter). This parameter can be used to measure how much a disk has
grown since it began forming stars.

Following the procedure described in Section~\ref{sSFRprof}, we
applied linear fits to our extinction-corrected $K$-band profiles to
obtain the central surface brightness $\mu_{K,0}$ (and the
corresponding central stellar mass surface density, $\Sigma_{M,0}$) as
well as the scale-length of the disk, which we consider to be the same
for both the $K$-band luminosity and the mass surface density profiles
(i.e., $\alpha_{K}=\alpha_{M}$) along with their errors. As explained
in the previous section, the assumption of constant $K$-band
mass-to-light ratio is a potential source of uncertainty for
$\alpha_{M}$. Again, considering the extreme case where
$\log(M/L_{K})$ decreases by 0.2\,dex in the outer regions of the
disk, $\alpha_{M}$ would be $\sim0.9\alpha_{K}$. For galaxies with the
opposite color gradient $\alpha_{M}$ could reach
$\sim1.1\alpha_{K}$. Therefore, our assumption that both
scales-lengths are equal might introduce a maximum systematic
uncertainty of only $\pm$10\%, depending on the sign of the radial
color gradient.

Monte Carlo simulations were carried out in order to properly derive
the uncertainties in $\alpha_{M}$, following the same methodology as
for the sSFR profiles. However, for some galaxies with low spatial
resolution and high photometric uncertainties (especially for the
outermost parts), upper and lower uncertainties for $\alpha_{M}$ were
extremely high (even greater than 100\%); those galaxies are marked
with a double dagger (\ddag) in Table~\ref{table3}.

Linear fits were only applied to points with $r>r_{in}$ located in the
disk-dominated region of the galaxy. In fact, a visual inspection of
the $K$-band profiles showed that our initial determination of
$r_{in}$ derived from the (FUV$-$$K$) color profiles was very
accurate, properly isolating the bulge-dominated part of the
galaxy. The value of $r_{in}$ was readjusted for only a few objects,
but always within our radial isophotal resolution (6\,arcsec). The
K-band surface brightness profiles can be seen in Fig.~3, along with
their corresponding fits. The fitting coefficients are shown in
Table~3. The best linear fits to the SFR profiles (within the same
radial ranges) have also been derived and the coefficients of these
fits are given in Table~3.

\subsection{Global statistical properties}\label{global}

Figure~4 shows histograms for both $\log(sSFR_{0})$ and
$m_{sSFR}$. Panel (a) shows that galaxies in our sample have values of
$sSFR_{0}$ ranging between over two orders of magnitude,
3$\times$10$^{-12}$\,yr$^{-1}$ and
3$\times$10$^{-10}$\,yr$^{-1}$. There is a clear dependency on the
morphological type: on average, Sc and later-type galaxies have
greater values of $sSFR_{0}$ than earlier types, although both
distributions overlap. Assuming that $sSFR_{0}$ is somehow related (at
least qualitatively) to the global sSFR, this result is consistent
with previous works based on total H$\alpha$ or UV photometry data,
where late-type spiral galaxies usually show higher specific star
formation rates than more massive early-type spirals (e.g$.$ Boselli
et al$.$ 2001; James et al$.$ 2004). This general behavior of the
total sSFR seems to be replicated by the extrapolated (central) value
of the sSFR of the disks. In other words: to some extent variations in
the y-intercept depend on changes of the global sSFR of the
disk. However, it must be emphasized that $sSFR_{0}$ is an
extrapolated value, and the integrated sSFR depends also on the slope
of the sSFR profile. Indeed, as we will show below, larger slopes are
usually found associated with lower y-intercepts ($sSFR_{0}$ values)
and vice versa. Therefore, all subsequent comparisons between
$sSFR_{0}$ and the global sSFR of the disk should be taken with care.

In panel (b) we show the histogram of the radial gradient of the
specific SFR. Most galaxies seem to have a slightly positive gradient
of sSFR. Although there are a few galaxies with negative values of
$m_{sSFR}$, the overall distribution favors positive sSFR
gradients. In addition, the histogram of early-type spirals ($T<5$)
seems to be more peaked or concentrated than the distribution of
late-type ones ($T\geq5$). As we will see later, this can be
understood in terms of the mass and size of each galaxy.

Figure~5a shows the specific SFR gradient as a function of
the y-intercept of the profiles. These parameters appear to be
correlated in the sense that galaxies with lower sSFR slopes tend to
have greater values of $\log(sSFR_{0})$ and vice versa. In other words,
we do not find many galaxies with simultaneously high (or low) values
of $m_{sSFR}$ and $\log(sSFR_{0})$. The observed trend is even tighter
if we limit ourselves to relatively large disks
($\alpha_{M}\geq3\mathrm{kpc}$), which lie within a narrow band in the
diagram; smaller galaxies have a greater dispersion.

The model-derived lines plotted in Fig.~5 allow us to understand the
relation between both parameters in terms of the size and total
specific SFR of the disks. Assuming exponential profiles for the
radial distributions of both SFR and stellar mass surface densities,
the specific SFR gradient can be expressed as (see Appendix B):
\begin{equation}
 m_{sSFR}=\left(\frac{1-\sqrt{\frac{sSFR_{0}}{sSFR}}}{\alpha_{M}}\right)\log e\label{m_y0}
\end{equation}
where $sSFR_{0}$ is the extrapolated specific SFR at $r=0$ and $sSFR$
is the total specific SFR of the disk (note that we are not including
the bulge here).

Therefore, Eq.~\ref{m_y0} provides the possible combinations of sSFR
at $r=0$ and its radial gradient which are compatible with a total
sSFR of the disk and a certain scale-length of the mass radial
profile. In Fig.~5a we have plotted six different curves
for disk-scales of 2, 4 and 8\,kpc and total specific SFR's of
$2\times10^{-11}\mathrm{yr}^{-1}$ and
$4\times10^{-10}\mathrm{yr}^{-1}$. We note that the general shape of
the distribution of the data can be well reproduced by Eq.~\ref{m_y0}
with a proper choice of reasonable values for these parameters. In
particular, the fact that for the same range in total sSFR
(2-40$\times$10$^{-11}$\,yr$^{-1}$) the curves for both small and
large disks nicely define the area of the diagram where the
corresponding galaxies are located indicates that physical size is the
main factor driving the differences and dispersion observed in the
values of the gradient of sSFR.

When analyzing the correlation between $m_{sSFR}$ and $\log(sSFR_{0})$
we must consider the possibility that this correlation could be partly
due to a degeneracy between these quantities. In order to determine
whether this is true or not, we have plotted the sets of simulated
values of both fitting parameters for each galaxy in Fig.~5b (in order
to avoid a complex graph, we only plot 200 out of the 2000 simulated
points for each galaxy). Each galaxy is represented by an elliptical
cloud of points that covers the region of the graph where the most
probable values of the fitting parameters are likely to be found. To
better appreciate the orientation and spatial coverage of these
clouds, we have used the covariance matrix to compute the ellipse that
contains 68\% of the simulated points for each galaxy. The colored
segments shown in the figure are the major axes of these ellipses; for
the sake of clarity, axes larger than 0.5 (in the units of the plot)
have been left out from the figure.

The confidence ellipses are found to be aligned in the same direction
(or slightly steeper) than the general distribution of data points,
with many of them clearly overlapping. Therefore, the observed global
correlation between the fitting parameters could be due $-$at least to
some extent$-$ to degeneracies between $m_{sSFR}$ and $\log(sSFR_{0})$
for each individual object. However, the major axes plotted in the
figure show that the confidence regions for many galaxies are small
enough to be considered detached one from another over the whole
ranges of $m_{sSFR}$ and $\log(sSFR_{0})$. Consequently, the global
shape of the correlation between these two parameters cannot be
explained just on the basis of individually correlated errors; there
do exist physical reasons that determine whether a certain combination
of slope and y-intercept of a sSFR profile is plausible or not.

\subsection{Dependency on size and mass}\label{trumpet}

In this and the following sections we analyze how both fitting
parameters, $\log(sSFR_{0})$ and $m_{sSFR}$ depend on several physical
parameters. In Fig.~6a we show the variation of the specific SFR
gradient with $\alpha_{M}$, the mass surface density scale-length of
the disk. The sample has been divided into three bins of $M_{K}$, in
order to simultaneously study the influence of the total mass. For the
sake of clarity, galaxies with error $\Delta\alpha_{M}>1\mathrm{ kpc}$
are shown without error bars (open symbols). Also, the ranges in both
axes have been stretched and adjusted to show all but four galaxies,
which have even higher values of $\alpha_{M}$ or $m_{sSFR}$ (albeit
with larger uncertainties; see Table~3). Their positions along the x/y
axis are marked by horizontal/vertical arrows, whose colors indicate
the corresponding $M_{K}$ bin. The black solid line at the background
indicates the average value of $m_{sSFR}$ in bins of 1.5 kpc, and the
gray shaded band corresponds to the 1$\sigma$ deviation with respect
to the mean. Both quantities, mean value and standard deviation, have
been computed using the whole set of 2000 values resulting from the
Monte Carlo simulations for each galaxy in its corresponding bin of
$\alpha_{M}$.

We can clearly see that less massive galaxies present quite different
values of $m_{sSFR}$, mostly positive, but also some negative. This
wide range of values, however, shrinks as we move towards larger and
more massive galaxies. At the high-mass end of the distribution, most
of the data points seem to concentrate within a relatively narrow
range, roughly centered around zero or slightly positive values of
$m_{sSFR}$ (note that a larger sample of big, massive galaxies would
be desirable to better constrain this asymptotic value). These results
are consistent with those derived by Taylor et al$.$ (2005) from the
analysis of (U$-$R) color profiles (uncorrected for internal
extinction) for a sample of 142 spiral, irregular and peculiar
galaxies. Small galaxies are indeed expected to exhibit a wider
variety of behaviors than larger ones, since the effects of the
spatial and temporal distribution of star formation episodes are $-$in
relative terms$-$ greater for them. Massive galaxies ought not be so
sensitive to external factors that could affect their star formation
histories (e.g$.$ gas accretion from a low-mass neighbor galaxy, ram
pressure stripping, etc$.$).

It could be argued that the higher dispersion observed in smaller and
less luminous galaxies could be just due to greater uncertainties in
$m_{sSFR}$. It is true that most galaxies with high values of $\Delta
m_{sSFR}$ lie in the low-size region of the diagram, typically below
2\,kpc, where the overall disperion is higher. However, if we plot
only the 122 galaxies for which $\Delta m_{sSFR}\leq0.04$ (a
representative value for galaxies all over our ranges of mass and
size), the trumpet-like shape of the diagram is preserved.

Panel (b) shows a similar graph, but this time with
$\log(sSFR_{0})$. Despite the dispersion of the data, we can see a
trend with mass and size, already hinted by the histogram in Fig.~4a:
small and less massive spiral galaxies (usually late-type ones) have
higher values of $sSFR_{0}$ (roughly between 0.5 and
1.5$\times$10$^{-11}$\,yr$^{-1}$) and then decreases as we consider
larger and more massive galaxies, although the dispersion is high and
the difference is mainly seen only between the two extreme bins of
$M_{K}$. This trend could be enhanced by $\sim 0.4$\,dex if, as
discussed in Section~\ref{sSFRprof}, the mass-to-light ratio varies
with Hubble type. Again, if we consider $sSFR_{0}$ to be a measure of
the overall level of the specific SFR, then it is not surprising that
its trend with size and mass is similar to the one exhibited by the
total sSFR deduced from global photometry data.

\subsection{Dependency on size and environment}\label{sigma5}
In this section we analyze the possible dependency of the fitting
parameters on size and environment. Panels (c) and (d) in Fig.~6 are
similar to (a) and (b), but the color scheme accounts for different
local galaxy densities, computed according to the methodology
described in Balogh et al$.$ (2004). For each galaxy we determine
$d_{5}$, the projected distance to the fifth neighbor that is brighter
than $M_{J}=-22$\,mag. We compute $\Sigma_{5}$, the projected local
density as being the number of galaxies within a circular area of
radius $r=d5$ and a redshift slice of $\pm1000$~km/s (in order to take
into account peculiar velocities). That is $\Sigma_{5}=N/(\pi
d^{2})$. Our magnitude limit of $M_{J}=-22$\,mag for the fifth
neighbor is consistent with the one used by Balogh et al$.$,
$M_{r}=-20$\,mag, assuming that $(r-J)\sim 2$ for typical spirals
(Peletier \& Balcells 1996; Fukugita, Shimasaku \& Ichikawa 1995).

We retrieved the coordinates and redshifts of neighbor galaxies for
each object in our sample from NED. Their magnitudes in both $J$ and
$K$ bands were collected from the 2MASS XSC catalog. It should be
noted that since neighbor galaxies were compiled from the different
surveys and sources provided by NED, our determinations of
$\Sigma_{5}$ are far from being uniform throughout the whole
sample. There may exist biases due to the different spatial coverage
of each survey. Besides, many galaxies in our sample are so nearby
that the search radius had to be extended up to the limit allowed by
NED (300 arcsec) in order to find enough neighbors. Due to that
limitation we could only compute $\Sigma_{5}$ for 74 galaxies (45\% of
the sample).

Panel (c) shows how the specific SFR gradient changes with size, with
galaxies sorted out into three bins of projected local density, in
units of Mpc$^{-2}$. Galaxies belonging to the Virgo cluster are
represented by black diamonds. There does not seem to exist any kind
of relation between $m_{sSFR}$ and $\Sigma_{5}$, although a larger
number of data points and more robust values of $\Sigma_{5}$ would be
desirable in order to confirm this. On the other hand, note that among
the six Virgo galaxies found in our sample, five of them have
$m_{sSFR}\lesssim 0$ and only one has $m_{sSFR}>0$, whereas nearly
70\% of the whole sample have positive sSFR gradients.

In panel (d) we carry out a similar study of the extrapolated sSFR at
the center of each galaxy. The dispersion of data is too high to
derive conclusive results. It is interesting to note that half of the
galaxies belonging to the Virgo cluster have $sSFR{0} \lesssim
10^{-11} yr^{-1}$, whereas this fraction drops to $\sim 13\%$ when we
consider all 74 galaxies in that plot. In any case, the reader is
cautioned that the results in panels (c) and (d) might be marginal
considering the high dispersion of the data.

The previous way of measuring the local galaxy density is global in
nature as it does not explicitly take into account the possible
interactions with the closest neighbors, which could play an important
role in the radial distribution of the sSFR. In order to study this
aspect of environment, for each galaxy we have computed the projected
distance to the nearest neighbor whose mass is at least 0.2 times the
mass of the galaxy itself (i.e$.$ being no fainter than 1.75\,mag
different in $K$). The color scheme in Figure~6e encodes different
distances to these neighbors, whereas several symbol sizes are used to
show the different mass ratios. The black curve shows the maximum sSFR
gradient expected for a given $\alpha_{M}$ according to a linear
disk-growth model with $\tau=\infty$ (see Sections~\ref{model} and
\ref{discussion}). No blue circles are seen below $\alpha_{M} \sim 2
\mathrm{\ kpc}$, since it is easier to find close neighbors over a
certain relative mass-ratio for the smallest and least massive
galaxies. No evident segregation is seen in $m_{sSFR}$, nor in
$sSFR_{0}$ [panel (f)].

\section{Modeling the specific SFR radial profiles}\label{model}
We have seen in previous sections that the specific SFR radial
gradient exhibits an interesting behavior: while there is a wide range
of observed values (both positive and negative) for galaxies with
disk-scales typically smaller than 2 or 3 kpc, this amplitude
decreases when we focus on increasingly larger disks, whose sSFR
slopes are generally very close to zero or only slightly positive.

In this section we try to reproduce this trumpet-like shape with a
relatively simple model of the radial and temporal evolution of the
star formation rate in these galaxies. According to previous work the
evolution of the `thin-disk' is thought to dominate the inside-out
growth of spiral galaxies (Chiappini, Matteucci, \& Gratton, 1997),
which is believed to start developing at $z\sim$1 (Brook et al$.$
2006). After this epoch mergers gradually become less intense and less
frequent. We may therefore suppose that since $z=1$ the growth of
spirals has been mainly driven by gradual star formation processes
taking place in their thin-disks.

We assume that the SFR density can be approximately modeled as:
\begin{equation}
 \Sigma_{SFR}(r,t)=\Sigma_{SFR}(0,0)\ e^{-t/\tau}e^{-r/(\alpha_{0}+bt)}\label{SFRrt}
\end{equation}
where we have set our temporal origin $t=0$ at
$z=1$. $\Sigma_{SFR}(0,0)$ is the central SFR surface density at $t=0$
(we do not make any hypothesis about its possible values since, as we
will see shortly, it vanishes when computing the specific SFR). In
Eq.~\ref{SFRrt} we have taken into account both temporal and radial
variations of the star formation rate. On the one hand, the overall
SFR is modulated by the global timescale, $\tau$, which should be of
the order of the gas-infall timescale divided by the index of the star
formation law. On the other hand, we have parametrized the
scale-length of the SFR profile as
$\alpha_{SFR}(t)=\alpha_{0}+bt$. Positive values of $b$ correspond to
disks in which the star formation is taking place in progressively
outward in the disk as time goes by, whereas negative values could be
used to describe SFR radial distributions whose extent decreases with
time.

We can compute the \textit{current} total stellar mass and SFR surface
density profiles as follows:
\begin{eqnarray}
 \Sigma_{M}(r,T)&=&(1-R)\int_{0}^{T} \Sigma_{SFR}(0,0)\ e^{-t/\tau}e^{-r/(\alpha_{0}+bt)} dt\\\label{massT}
\Sigma_{SFR}(r,T)&=&\Sigma_{SFR}(0,0)\ e^{-T/\tau}e^{-r/(\alpha_{0}+bT)}\label{SFRT}
\end{eqnarray}
where $T=7.72$ Gyr is the look-back time for $z=1$ and R is the
fraction of gas which is returned into the ISM. Dividing both
equations we obtain the present-day specific SFR profiles, from which
their radial gradients can be derived (see appendix C for
details). The exact value of R will not affect our sSFR gradients as
long as it does not change across the radius of the disk. This is
actually the case under the \textit{Instantaneous Recycling
Approximation} (IRA), which assumes that all stars with masses greater
than $1M_{\sun}$ die immediately, whereas the rest live forever. Under
this assumption, the returned fraction R is an instantaneous parameter
which does not depend on the SFH, and hence will remain constant
across the extent of the disk.

In short, we can use this simple model to `predict' the current values
of the sSFR slopes ($m_{sSFR}$) and scale-length of the mass radial
distributions ($\alpha_{M}$) as a function of three basic parameters:
the SFR timescale ($\tau$), the initial scale-length of the SFR
profile at $z=1$ ($\alpha_{0}$), and its growth rate ($b$). We can
therefore check if the physical assumptions considered in this model
lead to the observed dependency between $m_{sSFR}$ and $\alpha_{M}$
presented in Fig.~6.

\section{Discussion}\label{discussion}
We now proceed to use the results of the simple model described above
to reproduce the general trends seen in Fig.~6. Figures~7a \& 7b show
different model predictions for several sets of parameters along with
the observed data points. The initial scale-length of the SFR profiles
remains constant along solid lines (with the corresponding values
shown below each curve in kpc). Similarly, dashed lines are curves
with constant values of $b$, the growth rate of $\alpha_{SFR}$. They
are marked with rotated labels, in units of kpc/Gyr. Although the
maximum value of $b$ shown in the figures is 0.1 kpc/Gyr, simulations
were carried out up to $b=1.5$ kpc/Gyr, but the corresponding curves
lie very close to one another, just slightly above the curve for
$b=0.1$ kpc/Gyr, and hence are not plotted for the sake of
clarity. The thick gray line marks the loci of disks that have grown
by 25\% since $z=1$ (i.e$.$ $\alpha_{M}(T)=1.25\alpha_{0}$), which is
the value found by Trujillo \& Pohlen (2005) from the study of
intermediate-redshift disk-galaxies in the UDF.

In panel (a) we present the results of the model assuming
$\tau=\infty$. As expected, positive values of the specific SFR slope
require positive values of $b$ (i.e$.$ galaxies in which the SFR
profile has been growing with time towards the outer regions of the
disk) whereas galaxies with negative values of $m_{sSFR}$
are those with a decreasing $\alpha_{SFR}(t)$. The model predicts that
galaxies with $\alpha_{M}\lesssim$1\,kpc are expected to present a
wide range of sSFR slopes, whereas bigger ones are constrained within
a narrower region of the plot, with values of $m_{sSFR}$ close to
zero. In fact, there is some degeneracy in that region of the plot,
since galaxies with very different values of $\alpha_{0}$ and the
growth-rate $b$ end up with similar current values of $\alpha_{M}$ and
$m_{sSFR}$. The solid gray line, corresponding to disks which have
grown by 25\%, nicely bisects the overall distribution of data points.

In spite of the degeneracy, there exists an upper limit for the
possible gradient at a given $\alpha_{M}$ that leaves out many
galaxies in the sample. The model predicts lower sSFR gradients than
are observed at a given $\alpha_{M}$. In Section~\ref{massprof} we
discussed the possible effects that radial changes in the
mass-to-light ratio could have on $\alpha_{M}$ and $m_{sSFR}$, and
argued that galaxies with a typical radial color distribution (bluer
in the outer regions) could have smaller scale-lengths (by a factor of
$\sim$10\%) and higher specific SFR gradients (although possibly not
larger than $\sim$0.02 dex/kpc). This would slightly displace data
points with $m_{sSFR}>0$ towards the upper-left zone of the plot, but
to a much lesser extent than needed to correspond to the model
results.

Panel (b) shows the model predictions for a timescale of the gas
infall of 2\,Gyr. Comparing this diagram with panel (a) we can see
that the `isocurves' are somewhat stretched towards the
upper-right. In other words: since we are now reducing the amount of
present-day gas-infall, galaxies are required to have higher specific
SFR radial gradients to achieve a given present-day scale-length. This
is the reason why the models shown in panel (b) predict greater values
of $m_{sSFR}$ for a given scale-length of the mass
distribution. However, such short SFR timescales are expected only in
elliptical and giant spiral galaxies, whereas for smaller spirals
$\tau$ values of $\simeq$7\,Gyr (for which our simple model yields
nearly equal results to those with $\tau=\infty$) are commonly
inferred (Gavazzi et al$.$ 2002). With our data we cannot completely
discard the possibility of low values of $\tau$ for at least some
galaxies; however, it is worth noting that the average growth of 25\%
found by Trujillo \& Pohlen (2005) [gray line] is now a poorer average
value of the whole distribution. Hence, although introducing a short
SFR timescale might be appropriate for galaxies with the largest
values of $\alpha_{M}$, it cannot solve the discrepancy between the
model results and the observed data for the fraction of small disks
with the highest sSFR slopes.

One possible explanation is that the use of a linear function for the
temporal variation of $\alpha_{SFR}(t)$ (Eq.~\ref{SFRrt}) lets the
stellar mass surface density profile expand at a very similar rate to
that of the SFR itself, since it inherits the growth-rate of
$\alpha_{SFR}(t)$ through Eq.~\ref{massT}. Consequently, only certain
combinations of the model parameters lead to mass profiles that grow
slow enough compared with the SFR profiles so as to yield positive
present-day sSFR gradients. We have run tests using other analytic
functions to describe the temporal evolution of the SFR scale-length
(not shown), such as an exponential function, but similar upper limits
for the sSFR slope were encountered. The same limit is found when
exploring different star formation histories, such as one ``a la
Sandage'', which consists of a delayed exponential function (Sandage
1986, Gavazzi et al$.$ 2002). In other words: by describing the growth
of the SFR radial profile with a smooth continuous function we are not
allowing the model to take into account possible recent events that
could have triggered new star-forming events in the outer regions of
the galaxy, which would alter the current SFR profile without
significantly modifying the mass distribution. Note that a similar
limit is also obtained when adopting an early epoch (earlier than
$z$=1) for the onset of the inside-out (or outside-in) formation of
the disks.

We now study the effects on our model of adopting a scale length for
the SFR that evolves rapidly with time.  As a first approximation to
the real problem, we can just multiply the value of $\alpha_{SFR}(T)$
(i.e$.$ at $z=0$) for a certain factor, without modifying the
corresponding scale-length of the mass profile. Figures~7c and 7d show
the results of the model using an `enhanced' SFR profile at $z=0$,
with $\alpha_{SFR\mathrm{\ enhanced}}=2\times\alpha_{SFR\mathrm{\
linear}}$, where `linear' refers to the original model. Such an
episode of enhanced inside-out growth accomodates the high $m_{sSFR}$
values obtained for some galaxies. However, we should point out that
Figures~6c and 6e show that neither local galaxy density nor the
presence of close neighbors seem to drive this enhanced inside-out
growth. Figure~6e shows that the properties of neighbors (mass and
distance to the galaxy) do not seem to change above the upper limit of
$m_{sSFR}$ predicted by the linear evolution model.

The opposite scenario is also possible: galaxies which have undergone
a long phase of `linear' disk growth since $z=1$ may have recently
lost some of the gas in their outermost parts (possibly stripped off
by a neighboring galaxy or by ram pressure stripping). In panels (e)
and (f) we show how the predictions of the model change if we multiply
the final value of $\alpha_{SFR}$ by a factor 0.7, leading to a
`depressed' current SFR in the galaxy's outermost regions. From these
graphs we can conclude that the currently negative sSFR gradients seen
in many galaxies could be explained either by a long-term reduction of
the SFR disk (negative $b$, panels (a) and (b)) or a recent inhibition
of the star formation in the outer zones of an otherwise
linearly-evolving galaxy (panels (e) and (f)).

We should point out that the smooth SFH used in our simple model is
just a first approximation to the real scenario in which the star
formation activity of galaxies presumably fluctuates during their
lifetime. Thus, it is very likely that the present-day SFR radial
distribution deviates from the one predicted by Eq.~\ref{SFRT}, so the
model upper limit should be interpreted just as an time-averaged
quantity. The current stellar mass surface density can be still
computed with Eq.~\ref{massT}, since it is a cumulative parameter and
the fluctuations are expected to get averaged after the
integration. But the SFR is more dependent on the particular time of
observation, thus increasing the dispersion in $m_{sSFR}$ if the SF
activity is currently `enhanced' (even over the `linear' model upper
limit) or depressed.

Finally, some galaxies like NGC~4736 present bright inner rings of
intense star formation that can lead to negative values of $m_{sSFR}$;
these kinds of (presumably) transitory events are not considered in
our model either but might lead to an increase in the dispersion in
the $m_{sSFR}$ values.

\section{Summary and conclusions}
We have obtained specific SFR radial profiles for a sample of 161
moderately face-on spiral galaxies selected from the GALEX Atlas of
Nearby Galaxies (Gil de Paz et al$.$ 2006). Combining the FUV profiles
presented in the Atlas with $K$-band profiles measured on 2MASS images
we obtained (FUV$-K$) color profiles, which were then corrected from
foreground Galactic extinction and internal one. For the latter we
made use of the radial extinction profiles derived by Boissier et
al$.$ (2006) from the ratio of total-infrared to FUV luminosity; for
those galaxies in our sample without available TIR-to-FUV profiles,
(FUV$-$NUV) color profiles were used to infer the internal extinction
through the IRX-$\beta$ relation. The uncertainties associated with
the use of the IRX-$\beta$ law were considered when computing the
errors in the extinction-corrected (FUV$-K$) color profiles. The sSFR
profiles were inferred from relation between SFR and FUV luminosity
given by Kennicutt (1998) and assuming a mass-to-light ratio
$M/L_{K}=0.8M_{\sun}/L_{\sun,K}$.

We characterize these sSFR radial profiles through their slopes and
y-intercepts, derived from the linear fit applied to each
profile. Both fitting parameters are not independent one from another,
yet their possible combinations can be physically constrained in terms
of the total sSFR of the disk and its scale-length.

The extrapolated sSFR at $r=0$ seems to follow (at least
qualitatively) the same trends with morphological type, mass and size
as the total sSFR obtained from UV and H$\alpha$ global photometry in
previous works.

As for the sSFR gradient ($m_{sSFR}$), a clear trend is seen with mass
and size, in the sense that whereas a large dispersion is found for
small galaxies, which present both positive and negative sSFR
gradients, this scatter becomes considerably reduced as we consider
larger and more massive galaxies, for which the sSFR gradient is
nearly flat or slightly positive, consistent with a moderate
inside-out scenario of disk formation. This behavior can be
reproduced to some extent by assuming a simple description of the star
formation history of disks in which the typical scale-length of the
radial distribution of SFR varies linearly with time. This simple
assumption seems to explain the progressively more constrained values
of $m_{sSFR}$ in increasingly larger disks.

This model predicts an upper limit for $m_{sSFR}$ for each given
scale-length of the mass profile, $\alpha_{M}$, since the growth-rate
of both the SFR and mass radial profiles are `coupled'. There is,
however, a subset of galaxies in our sample whose sSFR slopes lie
clearly above this upper boundary. This limit depends on the adopted
SFR timescale, $\tau$. Lower values of $\tau$ yield higher upper
limits for $m_{sSFR}$, since galaxies must exhibit greater sSFR
gradients to achieve a given current size if gas is depleted in
shorter timescales. However, values of $\tau\sim 2$~Gyr are usually
inferred for elliptical galaxies and the earliest and most massive
spirals, while late-type spirals present typically larger SFR
timescales ($\sim7$~Gyr). Although we must not simply discard such low
values of $\tau$, it should also be noted that empirical measurements
of the disk-growth since $z=1$ (Trujillo \& Pohlen 2005) are in better
agreement with our model if higher values of $\tau$ are used.

Recent deviations from the continuous growth of the SFR radial
distribution could account for the observed excess in the sSFR slopes
for these galaxies, with present-day scale-lengths of the SFR being
much larger than in the past. When studying the possible dependence of
$m_{sSFR}$ on the environment, however, no clear correlation is
found. Disks with very high sSFR gradients do not seem to exhibit
neither different local galaxy densities nor closer and more massive
neighbors than the rest of the galaxies in the sample. This could be
due either to lack of robustness in our estimators of the environment
properties or to the fact that interactions with surrounding galaxies
could actually modify the sSFR profiles in different ways, depending
on the geometry of the interaction, its timescale, etc. Temporal
fluctuations of the SFH (not necessarily related to environmental
properties) might also account for the observed dispersion of sSFR
gradients.

Disks with currently negative sSFR slopes can be modeled with a
decreasing scale-length of the SFR (outside-in formation), but other
scenarios are also feasible. Ram-pressure stripping or transitory
episodes of enhanced star formation in the inner parts of the disk
can lead to a currently smaller SFR scale-length than in the past;
recent reductions of 30\% or even lower are enough to yield negative
present sSFR gradients even for galaxies that had been evolving
inside-out since the formation of their thin disks.

\acknowledgments

JCMM acknowledges the receipt of a Formaci\'on del Profesorado
Universitario fellowship from the Spanish Ministerio de Educaci\'on y
Ciencia. JCMM, AGdP, JZ, JG are partially financed by the Spanish
Programa Nacional de Astronom\'{\i}a y Astrof\'{\i}sica under grant
AYA2003-01676. AGdP is also financed by the MAGPOP EU Marie Curie
Research Training Network. We thank the anonymous referee for his/her
comments that have significantly improved the paper. This publication
makes use of data products from 2MASS which is a joint project of the
University of Massachusetts and the Infrared Processing and Analysis
Center/California Institute of Technology, funded by the National
Aeronautics and Space Administration and the National Science
Foundation. We have made also use of the NASA/IPAC Extragalactic
Database (NED), which is operated by the Jet Propulsion Laboratory,
California Institute of Technology (Caltech) under contract with
NASA. Stimulating conversations with Dr$.$ A$.$ Arag\'on-Salamanca are
gratefully acknowledged.

{\it Facilities:} \facility{GALEX}, \facility{FLWO:2MASS},
\facility{CTIO:2MASS}

\appendix
\section{Deriving sSFR from (FUV$-K$)}
The SFR can be computed from the apparent magnitude in the FUV (in the
AB system) using the calibration given by Kennicutt (1998):
\begin{equation}
\log(SFR)(\mathrm{M_{\sun} yr}^{-1})=2\log d\mathrm{(pc)}-0.4FUV-9.216
\end{equation}
Similarly, the K-band surface brightness profiles can be converted
into stellar mass surface density profiles as follows:
\begin{equation}
\log(M/M_{\sun})=\log(M/L_{K})-0.4(K+5-5\log d\mathrm{(pc)}-3.33)
\end{equation}
where $M/L_{K}$ is the stellar mass-to-light ratio in solar units, and
3.33 is the absolute K-band magnitude of the Sun in the Vega system
(Worthey 1994). Therefore, combining the previous equations we can
derive the specific SFR:
\begin{equation}
\log(sSFR)(\mathrm{yr}^{-1})=-0.4(FUV-K)-8.548-\log(M/L_{K})
\end{equation}

\section{Deriving the \textit{m$_{sSFR}$ - sSFR$_{0}$} relation}
Since light profiles of disks can be approximately described with an
exponential law, SFR and stellar mass surface densities may be modeled
in the same way:
\begin{mathletters}
\begin{eqnarray}
\Sigma_{SFR}&=&\Sigma_{SFR,0}\ e^{-r/\alpha_{SFR}}\label{sfr}\\
\Sigma_{M}&=&\Sigma_{M,0}\ e^{-r/\alpha_{M}} \label{m}
\end{eqnarray}
\end{mathletters}
where $\Sigma_{SFR,0}$ and $\Sigma_{M,0}$ are the central SFR and mass
surface densities, and $\alpha_{SFR}$ and $\alpha_{M}$ are the length
scales of both distributions. The total sSFR of the disk can be
computed by integrating \ref{sfr} and \ref{m} from $r=0$ to $r=\infty$
to obtain the total SFR and mass and then dividing both
quantities. Strictly speaking, a different choice of the integration
limits would not affect the total sSFR, since the functional form of
\ref{sfr} and \ref{m} is the same. In fact, from dimensional
considerations alone it is evident that the total mass of the disk
$M_{disk}\propto\alpha_{M}^{2}\Sigma_{M,0}$, and similarly
$SFR_{disk}$ with the same proportionality factor. Therefore,
\begin{equation}
 sSFR\equiv\frac{SFR_{disk}}{M_{disk}}=\left(\frac{\alpha_{SFR}}{\alpha_{M}}\right)^{2}\ \frac{\Sigma_{SFR,0}}{\Sigma_{M,0}}=\left(\frac{\alpha_{SFR}}{\alpha_{M}}\right)^{2}sSFR_{0}\label{SFR_M}
\end{equation}
Dividing \ref{sfr} by \ref{m} we obtain the specific SFR as a function of $r$. From the resulting expression we can write $m_{sSFR}$ as a function of both scale lengths:
\begin{equation}
m_{sSFR}=\left(\frac{1}{\alpha_{M}}-\frac{1}{\alpha_{SFR}}\right)\log e=\left(\frac{1-\alpha_{M}/\alpha_{SFR}}{\alpha_{M}}\right)\log e\label{m_alpha}
\end{equation}
Combining eqs. \ref{SFR_M} and \ref{m_alpha} to eliminate $\alpha_{M}/\alpha_{SFR}$ we obtain:
\begin{equation}
 m_{sSFR}=\left(\frac{1-\sqrt{\frac{sSFR_{0}}{sSFR}}}{\alpha_{M}}\right)\log e
\end{equation}

\section{Mathematical details of the modeling of the sSFR profiles}
As explained in Section~\ref{model}, we compute the present-day
specific SFR as:
\begin{equation}
sSFR(r,T)=\frac{\Sigma_{SFR}(r,T)}{\Sigma_{M}(r,T)}=\frac{e^{-T/\tau}e^{-r/(\alpha_{0}+bT)}}{(1-R)\int_{0}^{T} e^{-t/\tau}e^{-r/(\alpha_{0}+bt)} dt}
\end{equation}

The resulting expression for $sSFR(r,T)$ cannot be expressed in a
simple analytical form, and first order approximations lead to
oversimplified results, where both $\alpha_{M}$ and $\alpha_{SFR}$
expand at the same rate, that is, $\dot{\alpha}_{M}(t)=b$ and hence
$m_{sSFR}=0$. Therefore, we opted to derive the sSFR profiles
numerically, computing the integral in the previous expression using
the Gauss-Legendre quadrature algorithm:
\begin{equation}
 \int_{a}^{b}f(x)dx=\frac{b-a}{2}\int_{-1}^{1}f\left(\frac{b-a}{2}\xi+\frac{b+a}{2}\right)d\xi\simeq\frac{b-a}{2}\sum_{k=1}^{n}w(\xi_{k})f\left(\frac{b-a}{2}\xi+\frac{b+a}{2}\right)
\end{equation}
where the abscissas $\xi_{k}$ and their corresponding weights
$w(\xi_{k})$ can be derived from the Legendre polynomial
$P_{n}(x)$. Due to the low computational cost but high accuracy of
this method we decided to use $n=10$.

The resulting $\Sigma_{M}(r,T)$ and $sSFR(r,T)$ profiles slightly
deviate from an exponential law, but $\log(sSFR(r,T))$ and
$\log(\Sigma_{M}(r,T))$ can still be properly described by a straight
line. Since in principle $\alpha_{M}$ could depend on $r$, we obtain
an initial guess on $\alpha_{M}$ by fitting our model profiles between
$r=0$ and $r=50$ kpc. All subsequent fits needed to obtain the final
values for $\alpha_{M}$ and $m_{sSFR}$ are performed between
1.5$\times\alpha_{M}$ and 4.0$\times\alpha_{M}$. These limits were
those used in the numerical N-body simulations by Brook et al$.$
(2006) and are consistent with the values of $r_{in}$ (average value
0.8$\times$$\alpha_{M}$; see Section~\ref{sSFRprof} and Table~3) and
the outermost radial data-point measured in the 2MASS $K$-band images
of the galaxies in our sample (average value 4.4$\times$$\alpha_{M}$;
see Table~2). For $\tau=\infty$ the initial guess on $\alpha_{M}$
differs from the finally adopted value by less than 10\% for 93\% of
the simulated profiles, with the difference being less than 20\% for
the rest.

\clearpage
% Consulta en MySQL:
% select galex_reducida.object, RAhms, DECdms, format(i,0), galex_sample.E_B_V, galex_ned.mphtype, galex_ned.T, galex_ned.eT, galex_reducida.d, FUV_asympt, eFUV_asympt, format(K,2), format(eK,2), FUV_r50 from galex_reducida, galex_sample, uv_mags, galex_ned where select_flag='T' and galex_reducida.object=galex_sample.object and galex_reducida.object=uv_mags.object and galex_reducida.object=galex_ned.object order by galex_reducida.RA2000;
\clearpage
\begin{deluxetable}{lccrcccrrr}
\tabletypesize{\scriptsize}
\tablecolumns{10}
\tablecaption{Sample\label{table1}}
\tablewidth{0pt}
\tablehead{
\colhead{Object name} & \colhead{RA$_{2000}$} & \colhead{DEC$_{2000}$} & \colhead{D25} & \colhead{$i$} & \colhead{E(B$-$V)} & \colhead{$T$} & \colhead{dist} & \colhead{m$_{FUV}$} & \colhead{m$_{K}$} \\
\colhead{} & \colhead{(h:m:s)} & \colhead{(d:m:s)} & \colhead{(arcmin)} & \colhead{(deg)} & \colhead{(mag)} & \colhead{type} & \colhead{(Mpc)} & \colhead{(mag)} & \colhead{(mag)}\\
\colhead{(1)} & \colhead{(2)} & \colhead{(3)} & \colhead{(4)} & \colhead{(5)} & \colhead{(6)} & \colhead{(7)} & \colhead{(8)} & \colhead{(9)} & \colhead{(10)}}
\startdata
PGC~00282     & 00 04 01.5 &  $-$11 10 27.3 &    1.1 &      35 &  0.035 &    5 &    161   &  16.29$\pm$0.04 & 12.56$\pm$0.13  \\
ARP~256~NED02  & 00 18 50.1 &  $-$10 21 41.8 &    1.1 &      44 &  0.036 &  4.5 &   116   &  15.83$\pm$0.01 & 11.82$\pm$0.09  \\
NGC~0099      & 00 23 59.4 &    +15 46 13.5 &    1.4 &      22 &  0.056 &    6 &    77    &  15.13$\pm$0.01 & 11.76$\pm$0.10  \\
VV~548        & 00 36 02.0 &  $-$09 53 18.7 &    1.2 &      34 &  0.038 &    5 &    69    &  15.67$\pm$0.01 & 12.27$\pm$0.16  \\
NGC~0165      & 00 36 28.9 &  $-$10 06 22.2 &    1.5 &      30 &  0.034 &    4 &    83    &   16.6$\pm$0.01 & 10.37$\pm$0.07  \\
UGC~00372     & 00 37 27.9 &    +42 54 08.1 &    1.5 &      23 &  0.065 &    9 &    80    &  16.95$\pm$0.04 & 13.72$\pm$0.21  \\
PGC~02269     & 00 37 57.5 &  $-$09 15 09.3 &    1.4 &      31 &  0.035 &    6 &    74    &  15.64$\pm$0.02 & 11.60$\pm$0.12  \\
NGC~0195      & 00 39 35.8 &  $-$09 11 40.3 &    1.1 &      44 &  0.033 &    1 &    69    &  17.64$\pm$0.07 & 10.36$\pm$0.04  \\
\enddata
\tablecomments{Sample. (1): Galaxy name. (2): RA(J2000) of the galaxy center. (3): DEC(J2000) of the galaxy center. (4): Apparent major isophotal diameter at $\mu_{B}$=25 mag arcsec$^{-2}$ from the RC3 catalog. (5): Inclination angle, corrected for intrinsic thickness. \ddag Note that for NGC~1326 $q_{0}=0.4$ was assumed, since no value for $T=-1$ is available. (6): Galactic color excess from Schlegel et al$.$ (1998). (7): Morphological type T as given in the RC3 catalog. (8): Distance to the galaxy, rounded to the nearest Mpc when larger than 10 Mpc. (9): Apparent $FUV$ magnitude (AB system). (10): Apparent $K$-band magnitude (Vega system).}
\end{deluxetable}

\clearpage
\begin{deluxetable}{rrrrrrrr}
\tabletypesize{\scriptsize}
\tablecolumns{8}
\tablecaption{Radial profiles\label{table2}}
\tablewidth{0pt}
\tablehead{
\colhead{$a$} & \colhead{$a$} & \colhead{$\mu_{FUV}$} & \colhead{$\mu_{K}$} & \colhead{(FUV$-$K)$_{obs}$} & \colhead{A$_{FUV}$} & \colhead{(FUV$-$K)$_{corr}$} & \colhead{log(sSFR)} \\
\colhead{(arcsec)} & \colhead{(kpc)} & \colhead{(mag arcsec$^{-2}$)} & \colhead{(mag arcsec$^{-2}$)} & \colhead{(mag)} & \colhead{(mag)} & \colhead{(mag)} & \colhead{(10$^{-11}$ yr$^{-1}$)} \\
\colhead{(1)} & \colhead{(2)} & \colhead{(3)} & \colhead{(4)} & \colhead{(5)} & \colhead{(6)} & \colhead{(7)} & \colhead{(8)}}
\startdata
\cutinhead{PGC~00282}
6 &  4.69 & 24.64 $\pm$ 0.07 & 19.19 $\pm$ 0.09 & 5.45 $\pm$ 0.11 & 1.38 $^{+ 0.39 }_{- 0.36 }$ & 4.14 $^{+ 0.38 }_{- 0.40 }$ & 0.89 $^{+ 0.16 }_{- 0.15 }$ \\
12 &  9.38 & 24.37 $\pm$ 0.06 & 20.12 $\pm$ 0.16 & 4.25 $\pm$ 0.17 & 0.86 $^{+ 0.34 }_{- 0.31 }$ & 3.42 $^{+ 0.35 }_{- 0.38 }$ & 1.18 $^{+ 0.15 }_{- 0.14 }$ \\
18 & 14.07 & 24.35 $\pm$ 0.03 & 20.58 $\pm$ 0.24 & 3.77 $\pm$ 0.25 & 1.01 $^{+ 0.36 }_{- 0.32 }$ & 2.81 $^{+ 0.41 }_{- 0.43 }$ & 1.43 $^{+ 0.17 }_{- 0.16 }$ \\
24 & 18.76 & 25.06 $\pm$ 0.04 & 21.36 $\pm$ 0.40 & 3.70 $\pm$ 0.40 & 0.89 $^{+ 0.35 }_{- 0.31 }$ & 2.85 $^{+ 0.51 }_{- 0.53 }$ & 1.41 $^{+ 0.21 }_{- 0.20 }$ \\
\cutinhead{ARP~256~NED02}
6 &  3.37 & 22.91 $\pm$ 0.02 & 18.19 $\pm$ 0.10 & 4.72 $\pm$ 0.10 & 1.08 $^{+ 0.36 }_{- 0.33 }$ & 3.69 $^{+ 0.35 }_{- 0.38 }$ & 1.07 $^{+ 0.15 }_{- 0.14 }$ \\
12 &  6.75 & 23.76 $\pm$ 0.03 & 19.44 $\pm$ 0.15 & 4.32 $\pm$ 0.16 & 1.57 $^{+ 0.40 }_{- 0.38 }$ & 2.82 $^{+ 0.41 }_{- 0.43 }$ & 1.42 $^{+ 0.17 }_{- 0.16 }$ \\
18 & 10.12 & 25.08 $\pm$ 0.03 & 20.72 $\pm$ 0.37 & 4.36 $\pm$ 0.37 & 1.82 $^{+ 0.42 }_{- 0.40 }$ & 2.62 $^{+ 0.54 }_{- 0.56 }$ & 1.50 $^{+ 0.23 }_{- 0.22 }$ \\
24 & 13.49 & 25.19 $\pm$ 0.03 & 20.95 $\pm$ 0.41 & 4.24 $\pm$ 0.41 & 1.36 $^{+ 0.39 }_{- 0.36 }$ & 2.95 $^{+ 0.54 }_{- 0.56 }$ & 1.37 $^{+ 0.23 }_{- 0.22 }$ \\
\enddata
\tablecomments{Radial profiles. (1): Radius along semimajor axis in arcsec. (2): Radius along semimajor axis in kpc. (3): Observed surface brightness in FUV, corrected only from Galactic extinction. (4): Observed surface brightness in K, corrected only from Galactic extincion. (5): Observed (FUV$-$K) color, corrected only from galactic extinction. (6): Internal extinction in FUV. (7): (FUV$-$K) color corrected from both Galactic and internal extinction. (8): Logarithm of specific star formation rate (SFR per unit of stellar mass).}
\end{deluxetable}

% Consulta en MysQL:
% select object, r_bulge_arcsec+3 as rin, format(m,3) as m, format(m_e2,3) as me2, format(m_e1,3) as me1, format(y0,3) as y0, format(y0_e2,3) as y0e2, format(y0_e1,3) as y0e1, muM0, muM0_e2, muM0_e1, format(alfaK,2) as alfaK, format(alfaK_e2,2) as alfaKe2, format(alfaK_e1,2) as alfaKe1, format(mAFUV,3) as mAFUV, format(mAFUVe2,3) as mAFUVe2, format(mAFUVe1,3) as mAFUVe1, format(muSFR0*1e+8,2) as muSFR0,  format(muSFR0e2*1e+8,2) as muSFR0e2, format(muSFR0e1*1e+8,2) as muSFR0e1, format(alfaFUV,2) as alfaFUV, format(alfaFUVe2,2) as alfaFUVe2, format(alfaFUVe1,2) as alfaFUVe1 from galex_reducida where select_flag='T' order by RA2000;
\clearpage
\begin{deluxetable}{lrrrrrrrrr}
\tabletypesize{\scriptsize}
\tablecolumns{10}
\rotate
\tablecaption{Fit results\label{table3}}
\tablewidth{0pt}
\tablehead{
\colhead{Object name} & \colhead{$r_{in}$} & \colhead{$m_{sSFR}$} & \colhead{$\log(sSFR_{0})$} & \colhead{$\Sigma_{M,0}$} & \colhead{$\alpha_{M}$} & \colhead{$\Sigma_{SFR,0}$} & \colhead{$\alpha_{SFR}$} & \colhead{m$_{A_{FUV}}$} & \colhead{A$_{FUV,0}$}\\
\colhead{} & \colhead{(arcsec)} & \colhead{(dex/kpc)} & \colhead{($10^{-11} yr^{-1}$)} & \colhead{($M_{\sun}pc^{-2}$)} & \colhead{(kpc)}  & \colhead{($10^{-8} M_{\sun}yr^{-1}pc^{-2}$)} & \colhead{(kpc)} & \colhead{(mag/kpc)} & \colhead{(mag)}\\
\colhead{(1)} & \colhead{(2)} & \colhead{(3)} & \colhead{(4)} & \colhead{(5)} & \colhead{(6)} & \colhead{(7)} & \colhead{(8)} & \colhead{(9)} & \colhead{(10)}}
\startdata
PGC~00282     &    9 &    0.025$^{+0.028}_{-0.027}$ &    0.993   $^{+0.370}_{-0.369}$ &   223 $^{+  157}_{-    90}$ &  8.22\ddag                         &  2.20 $^{+  2.20}_{-   1.06}$ &     15.34\ddag                           &       0.003 $^{+   0.051}_{-   0.048}$ &    0.881 $^{+   0.731}_{-   0.715}$ \\
ARP~256~NED02 &    9 &    0.004$^{+0.034}_{-0.032}$ &    1.405   $^{+0.351}_{-0.356}$ &   473 $^{+  415}_{-   216}$ &  4.71\ddag                         &  12.01 $^{+ 7.96}_{-   4.65}$ &     4.91 $^{+   1.65}_{-     0.76     }$ &    $-$0.045 $^{+   0.052}_{-   0.048}$ &    2.024 $^{+   0.619}_{-   0.649}$ \\
NGC~0099      &    9 &    0.028$^{+0.048}_{-0.049}$ &    1.124   $^{+0.363}_{-0.348}$ &   571 $^{+  553}_{-   282}$ &  3.33\ddag                         &  7.59 $^{+  4.25}_{-   2.51}$ &     4.25 $^{+   1.32}_{-     0.66     }$ &    $-$0.059 $^{+   0.049}_{-   0.045}$ &    1.390 $^{+   0.452}_{-   0.472}$ \\
VV~548        &    9 & $-$0.006$^{+0.062}_{-0.063}$ &    1.475   $^{+0.386}_{-0.371}$ &   135 $^{+   86}_{-    53}$ &  5.89\ddag                         &  4.03 $^{+  3.85}_{-   1.92}$ &     5.42\ddag                            &    $-$0.012 $^{+   0.119}_{-   0.121}$ &    1.114 $^{+   0.789}_{-   0.723}$ \\
NGC~0165      &    9 &    0.001$^{+0.018}_{-0.018}$ &    0.671   $^{+0.199}_{-0.187}$ &   558 $^{+  217}_{-   155}$ &  5.22 $^{+  1.38}_{-    0.75  }$   &  2.62 $^{+  0.82}_{-   0.59}$ &     5.28 $^{+   0.58}_{-     0.42     }$ &       0.034 $^{+   0.021}_{-   0.022}$ &    1.165 $^{+   0.340}_{-   0.316}$ \\
UGC~00372     &    9 &    0.022$^{+0.062}_{-0.065}$ &    0.993   $^{+0.440}_{-0.398}$ &   151 $^{+  156}_{-    75}$ &  3.46\ddag                         &  1.49 $^{+  1.49}_{-   0.69}$ &     4.21\ddag                            &    $-$0.005 $^{+   0.104}_{-   0.099}$ &    0.893 $^{+   0.725}_{-   0.735}$ \\
\enddata
\tablecomments{Fit results. (1): Galaxy name. (2): Radius at which the contribution of the disk to the (FU$-$$K$) color begins to dominate over that of the bulge. (3): Specific SFR gradient, i.e.$\Delta\log(sSFR)/\Delta r$. (4): Logarithm of the extrapolated value of sSFR at r=0. (5): Stellar mass surface density extrapolated at r=0. Note that these values have not been corrected for inclination. (6): Scale-length of the stellar mass surface density profile. Galaxies marked with a double dagger (\ddag) are those for which $\Delta \alpha_{M}$ could not be properly computed (see text for details). (7): SFR surface density extrapolated at r=0. Note that these values have not been corrected for inclination. (8): Scale-length of the SFR surface density profile. Galaxies marked with a double dagger (\ddag) are those for which $\Delta \alpha_{SFR}$ could not be properly computed (see text for details). (9): Radial gradient of the extinction in the FUV. (10): Extrapolated value of the extinction in the FUV at $r=0$ (note that since it is an extrapolated value from the linear fits, negative values are possible, but without any physical meaning).}
\end{deluxetable}

\clearpage
\begin{figure}
\figurenum{1}
\epsscale{1.0}
\resizebox{1\hsize}{!}{\includegraphics*[125,260][485,635]{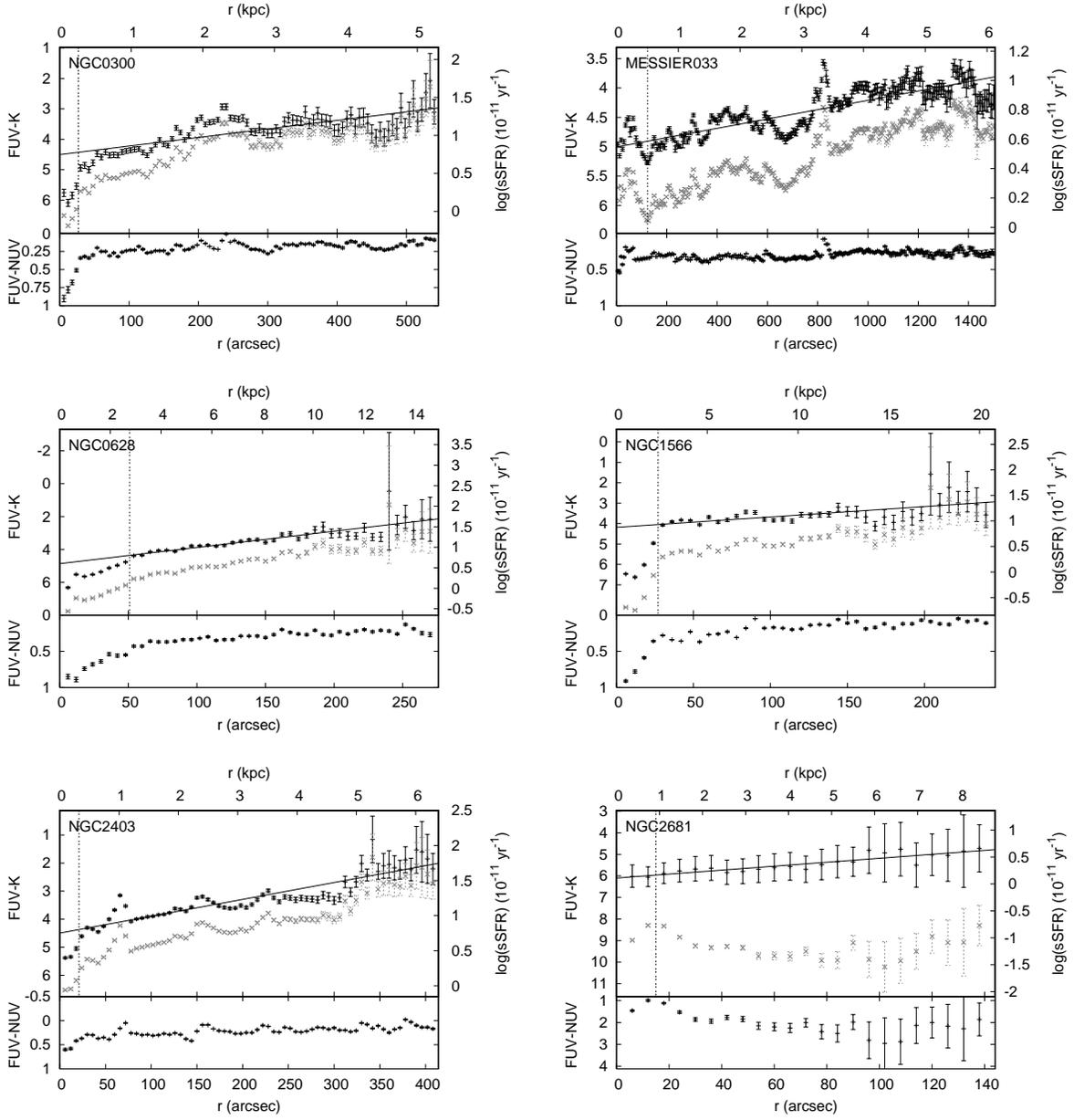}}
\caption{Sample galaxy profiles. Top: (FUV$-K$) color profiles, corrected only for foreground Galactic extinction (gray points) and for both Galactic and internal extinction (black points). Bottom: (FUV$-$NUV) color profiles. The vertical dotted line in each plot represents the radius at which the contribution of the bulge to the light profile becomes negligible compared to that of the disk. The solid line corresponds to the linear fit performed to the fully corrected data.}
\end{figure}

\begin{figure}
\figurenum{2}
\epsscale{1.0}
\resizebox{1\hsize}{!}{\includegraphics*{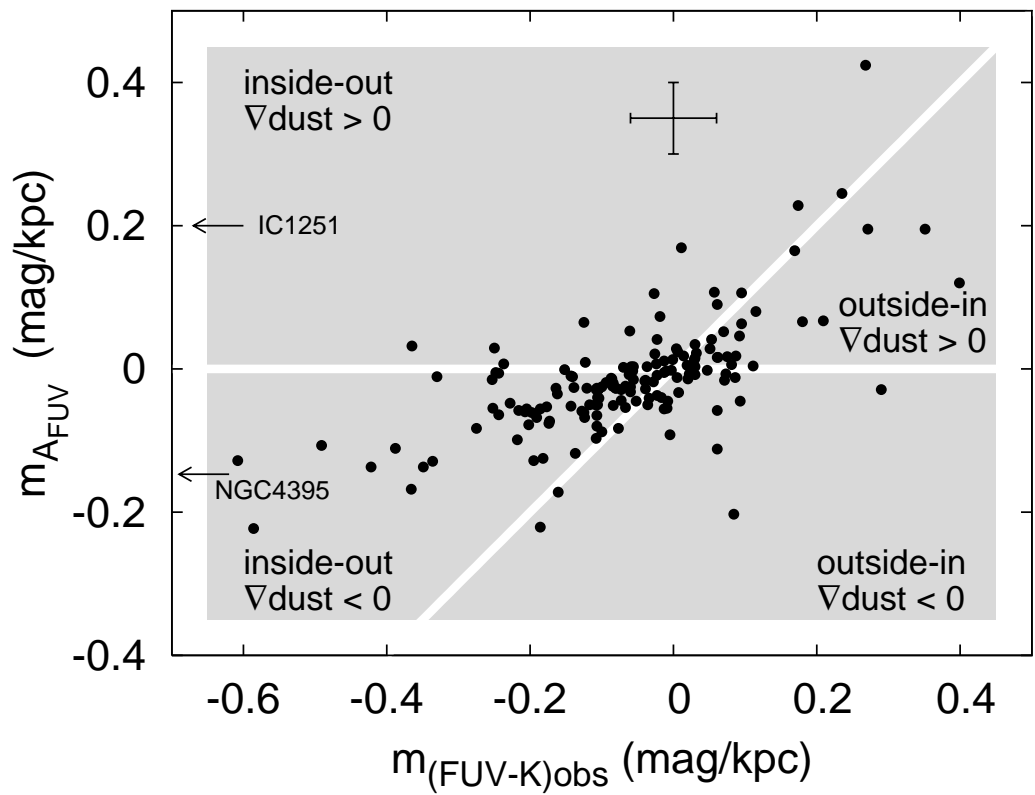}}
\caption{Radial gradient of the extinction in the FUV as a function of the observed gradient of (FUV$-K$) color. The cross shows the mean uncertainties in both parameters. The plot has been divided into four regions. The diagonal line corresponds to galaxies in which the observed color gradient is only due to radial variations of the dust content [i.e.: their intrinsic (FUV$-K$) profiles are flat]. Data points to the left of this line are then consistent with an inside-out formation of disks (and vice versa). The horizontal line sorts out the galaxies depending on whether the dust content decreases with radius (lower half of the plot) or increases (upper half). Galaxies out of range have been marked with arrows.}
\end{figure}

\clearpage
\begin{figure}
\figurenum{3}
\epsscale{1.0}
\resizebox{1\hsize}{!}{\includegraphics*[125,260][485,635]{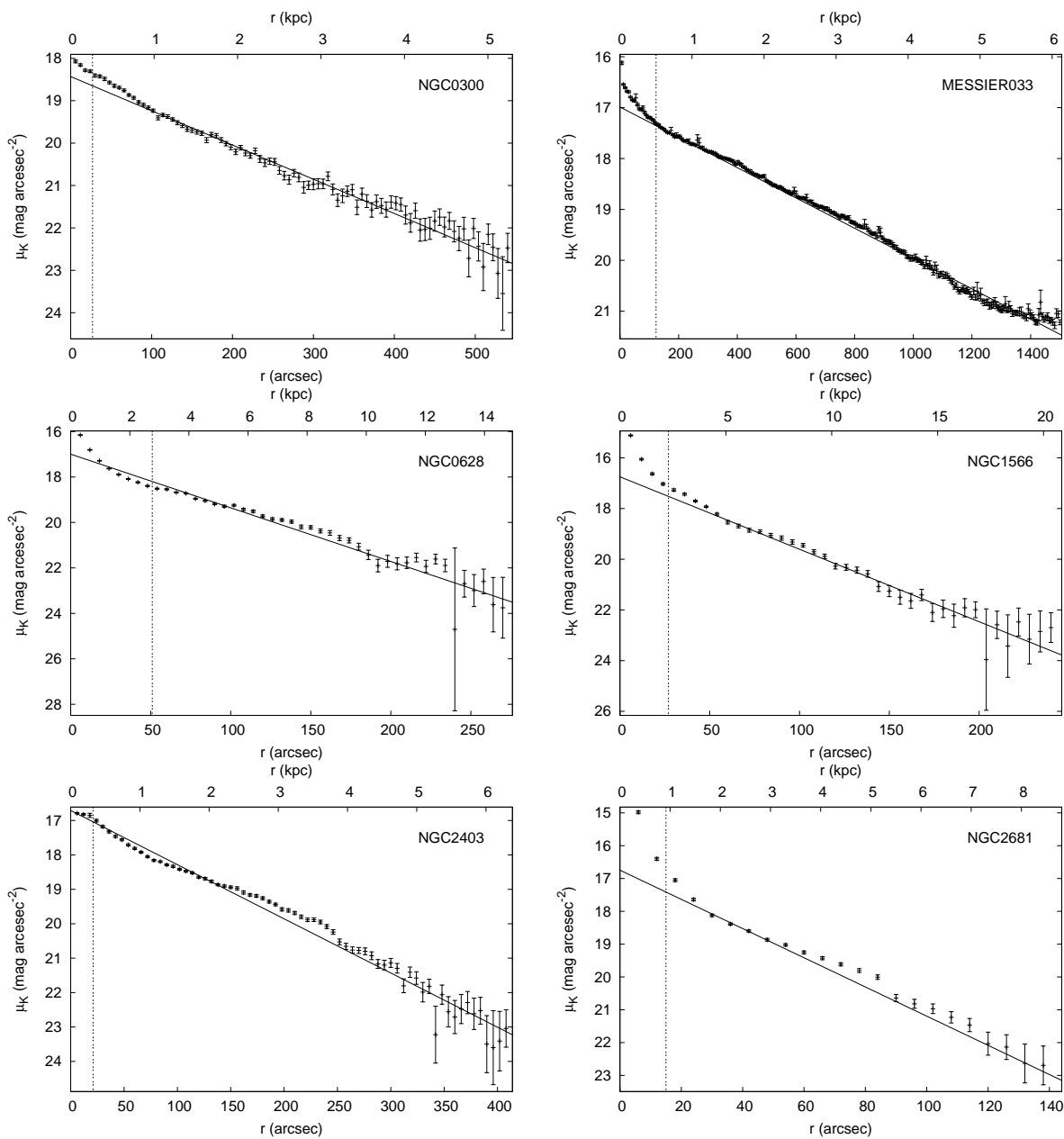}}
\caption{Sample K-band surface brightness profiles. The vertical dotted line in each plot represents the radius at which the contribution of the bulge to the light profile becomes negligible compared to that of the disk (same as in Fig.~1). The solid line corresponds to the linear fit performed to the profiles in the disk region.}
\end{figure}

\begin{figure}
\figurenum{4}
\epsscale{1.0}
\plottwo{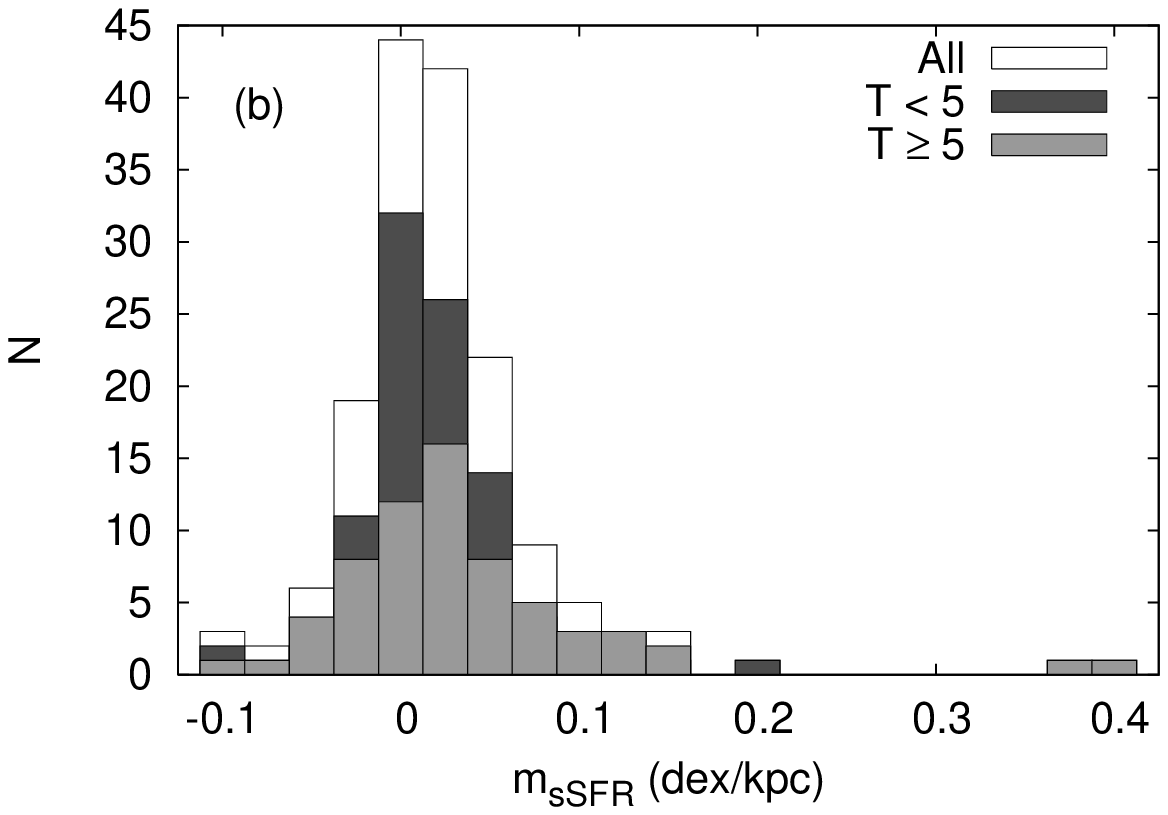}{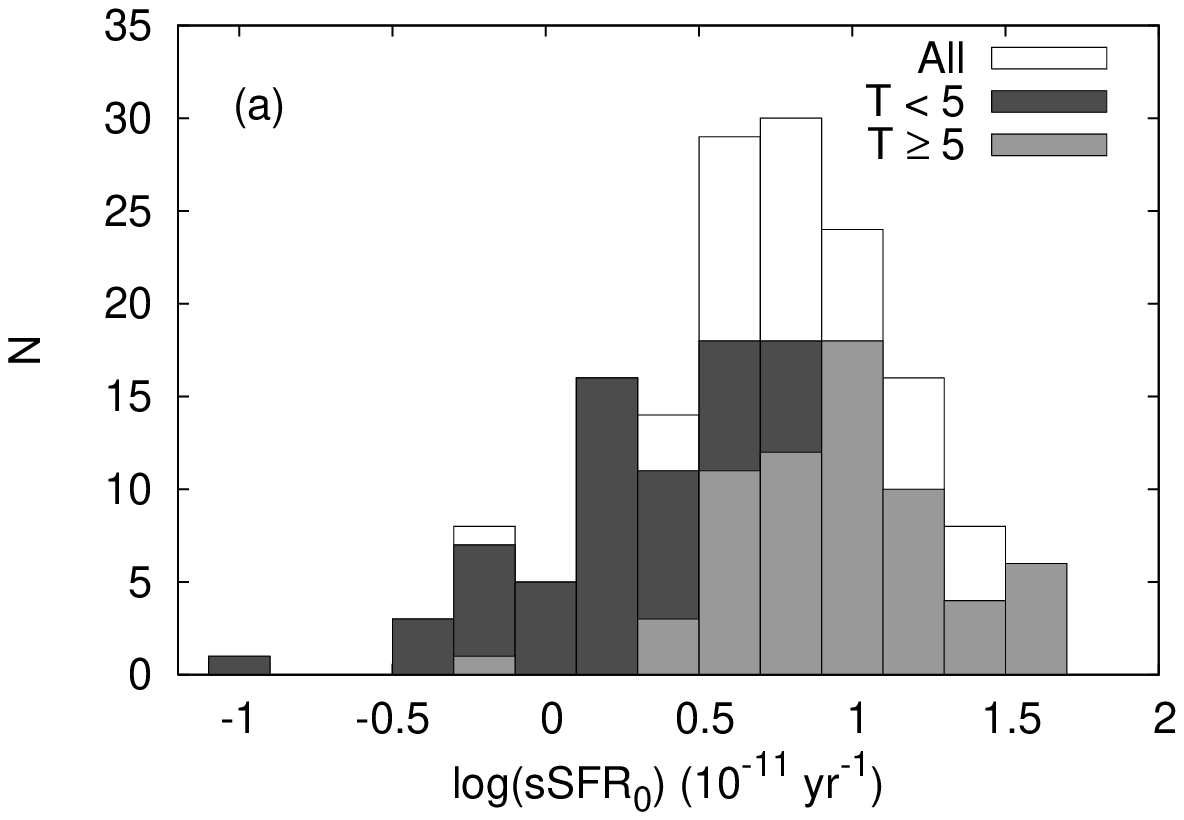}
\caption{(a) Histogram of the specific SFR extrapolated to $r$ = 0. (b) Histogram of the specific SFR gradient ($m_{sSFR}=\Delta \log(sSFR)/\Delta r$). Early ($T$$<$5) and late type ($T$$\geq$5) galaxies are distinguished by dark and light shading of the histograms respectively.}
\end{figure}

\clearpage
\centering{
\epsscale{1.0}
\resizebox{!}{0.48\vsize}{\includegraphics{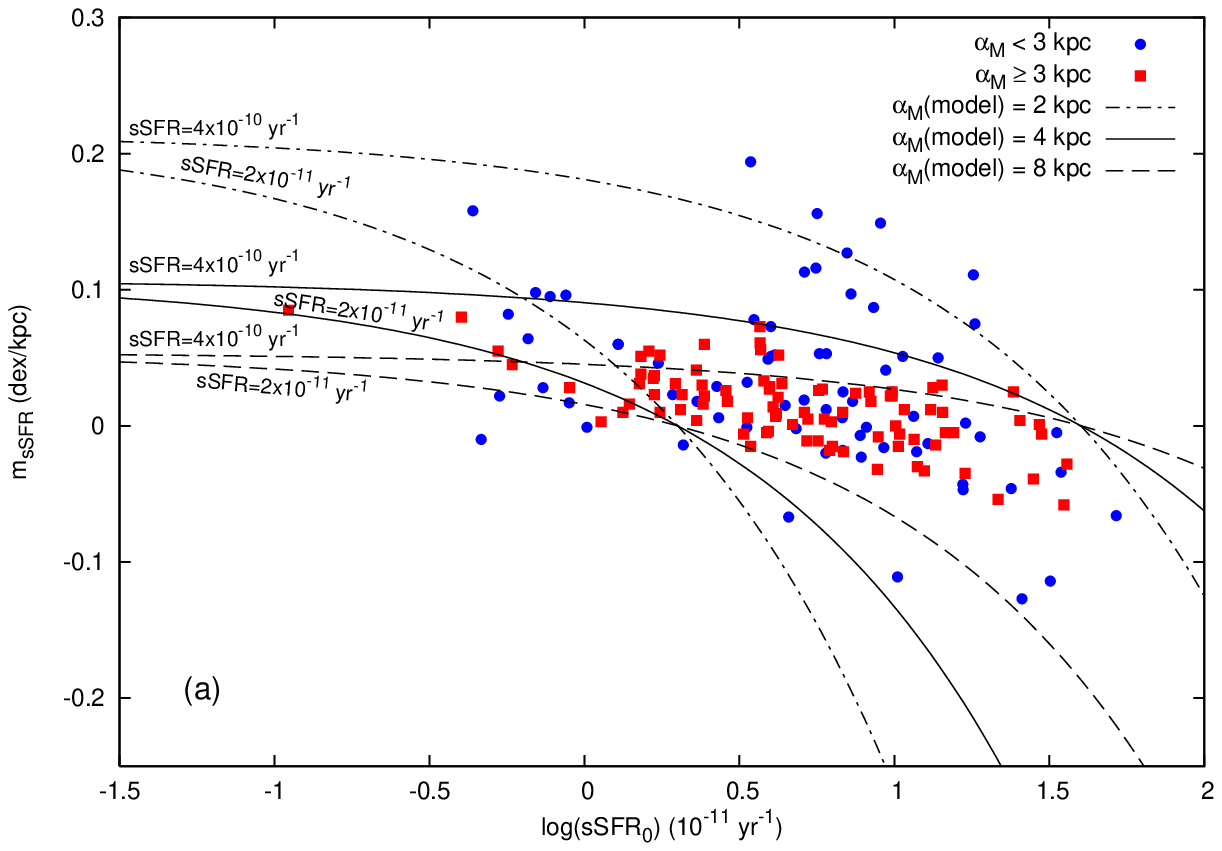}}\\
\vspace{0.5cm}
\resizebox{!}{0.48\vsize}{\includegraphics*[50,50][410,302]{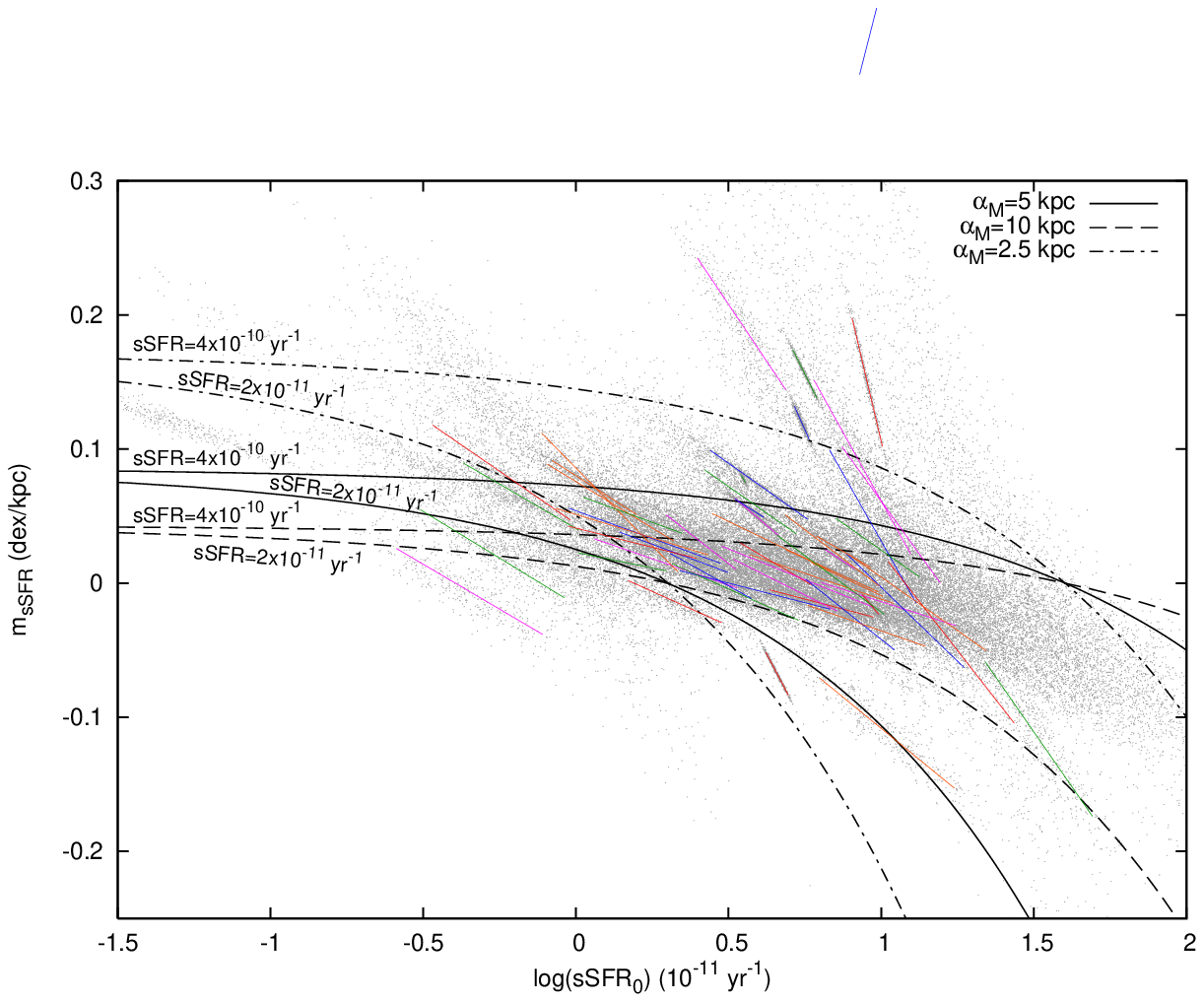}}\\
}

\clearpage
\begin{figure}
\figurenum{5}
\caption{(a) Relation between fitting parameters for early- and late-type spirals in our sample. Black curves show the theoretical relation between both parameters for a given total sSFR of the disk and a scale-length of the mass surface density profile. (b) Probability distributions of fitting parameters for all galaxies in the sample. Colored segments are the major axes of the 1-$\sigma$ confidence ellipses for each cloud of points (only for the 34\% of galaxies with the lowest uncertainties).}
\end{figure}

\clearpage
\centering{
\epsscale{1.0}
\resizebox{0.48\hsize}{!}{\includegraphics{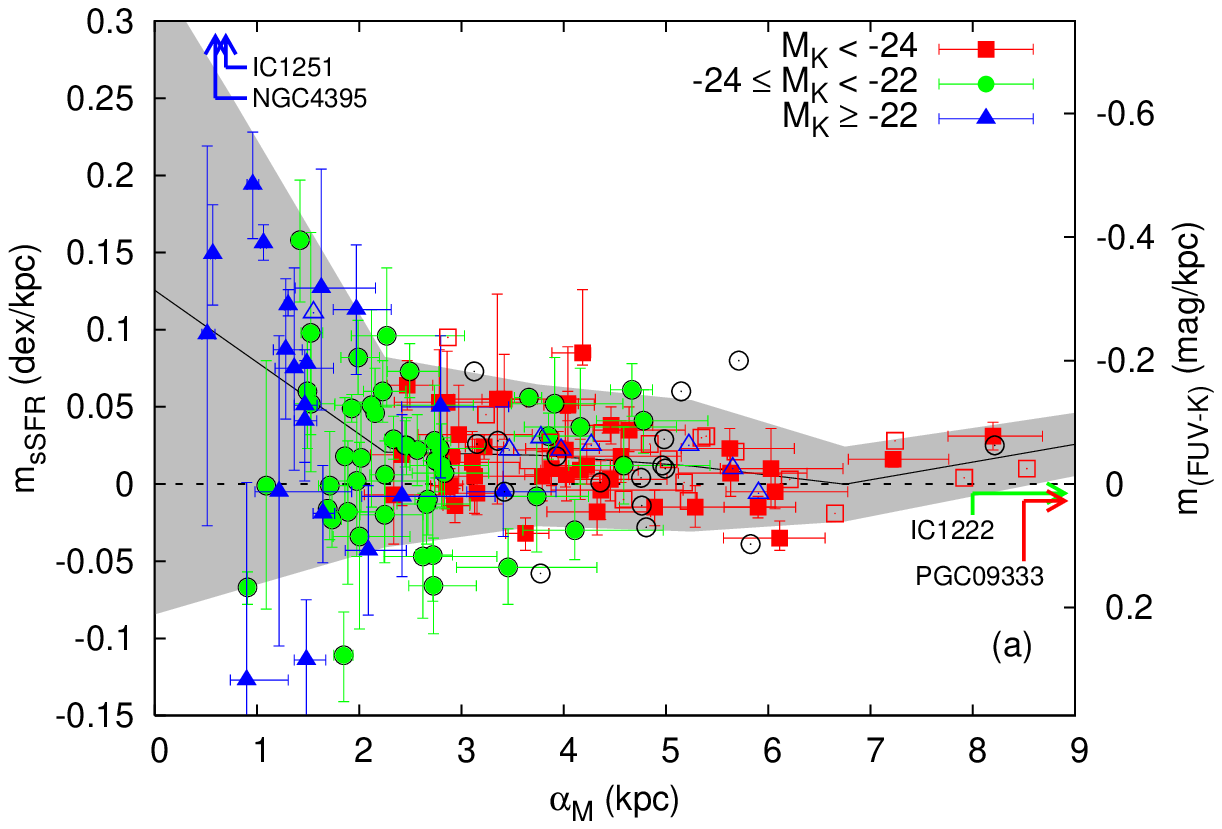}}\hspace{0.1cm}
\resizebox{0.48\hsize}{!}{\includegraphics{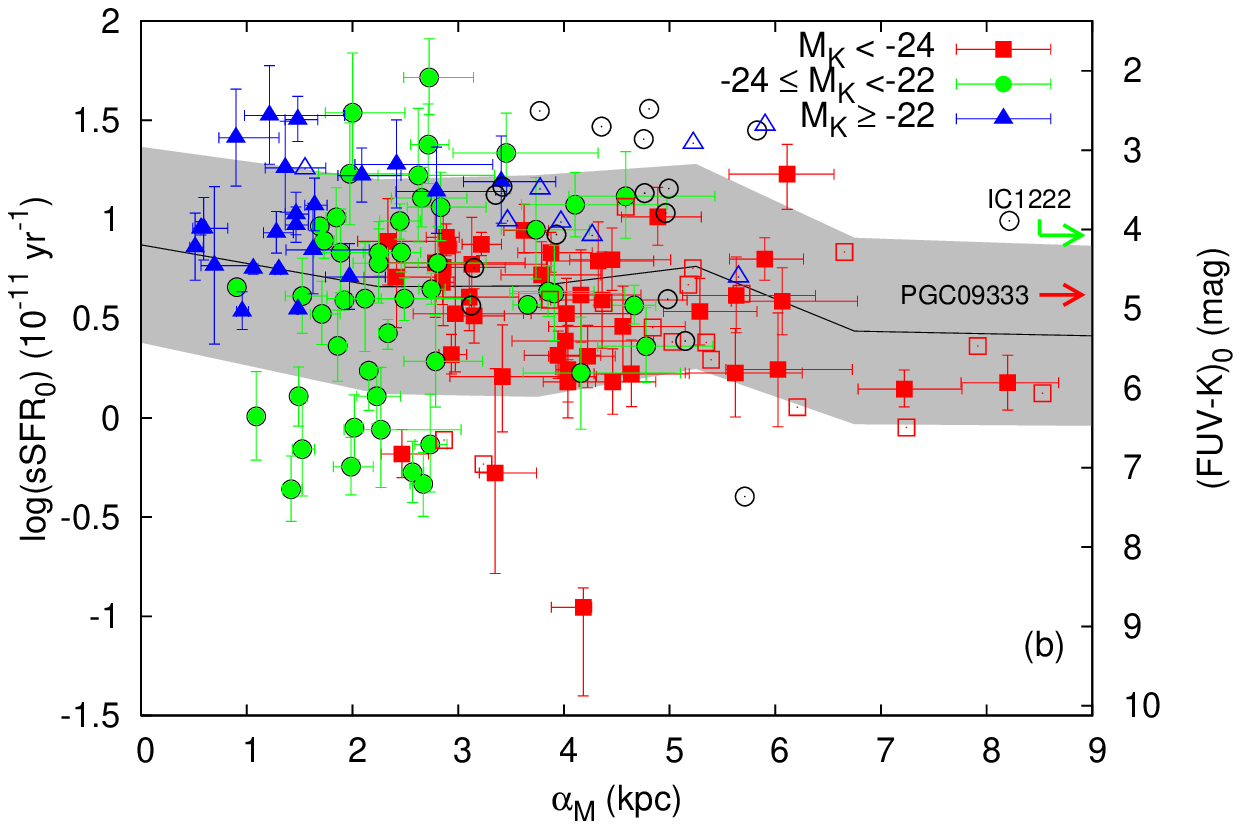}}\\
\resizebox{0.48\hsize}{!}{\includegraphics{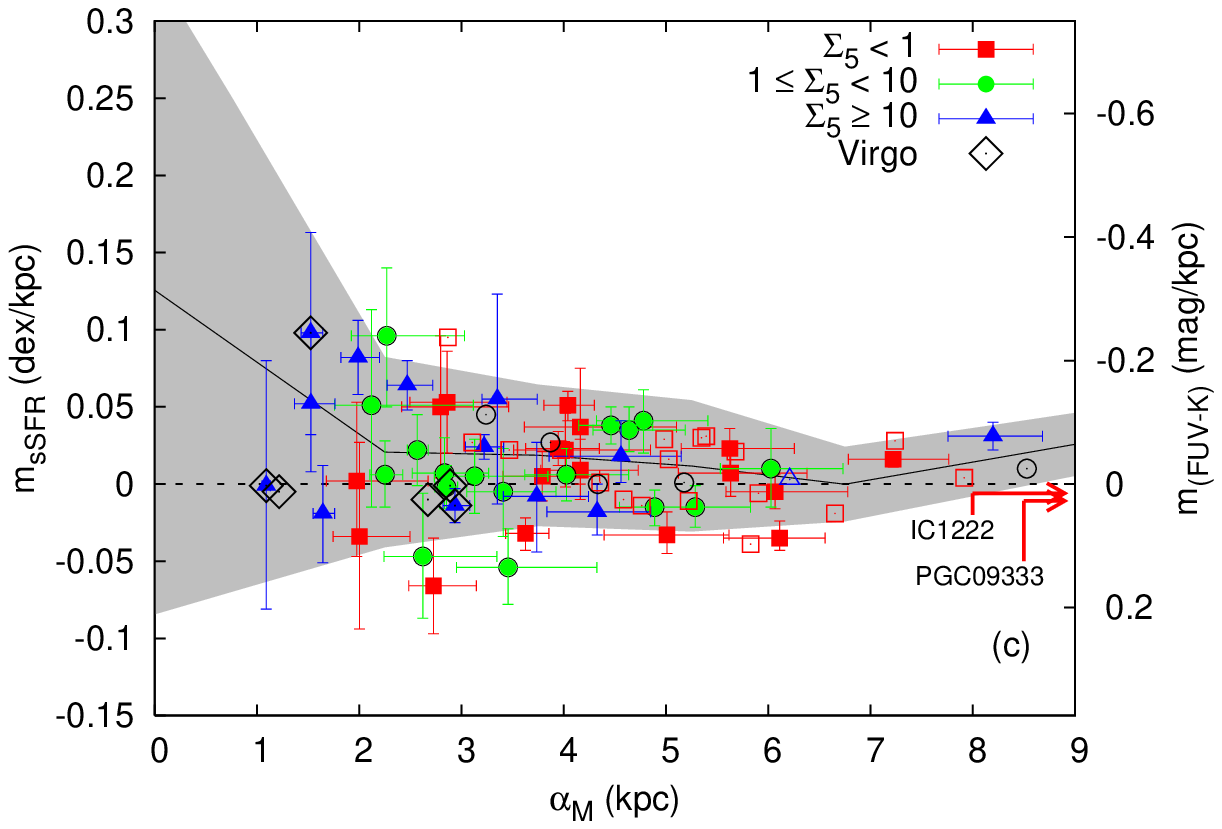}}\hspace{0.1cm}
\resizebox{0.48\hsize}{!}{\includegraphics{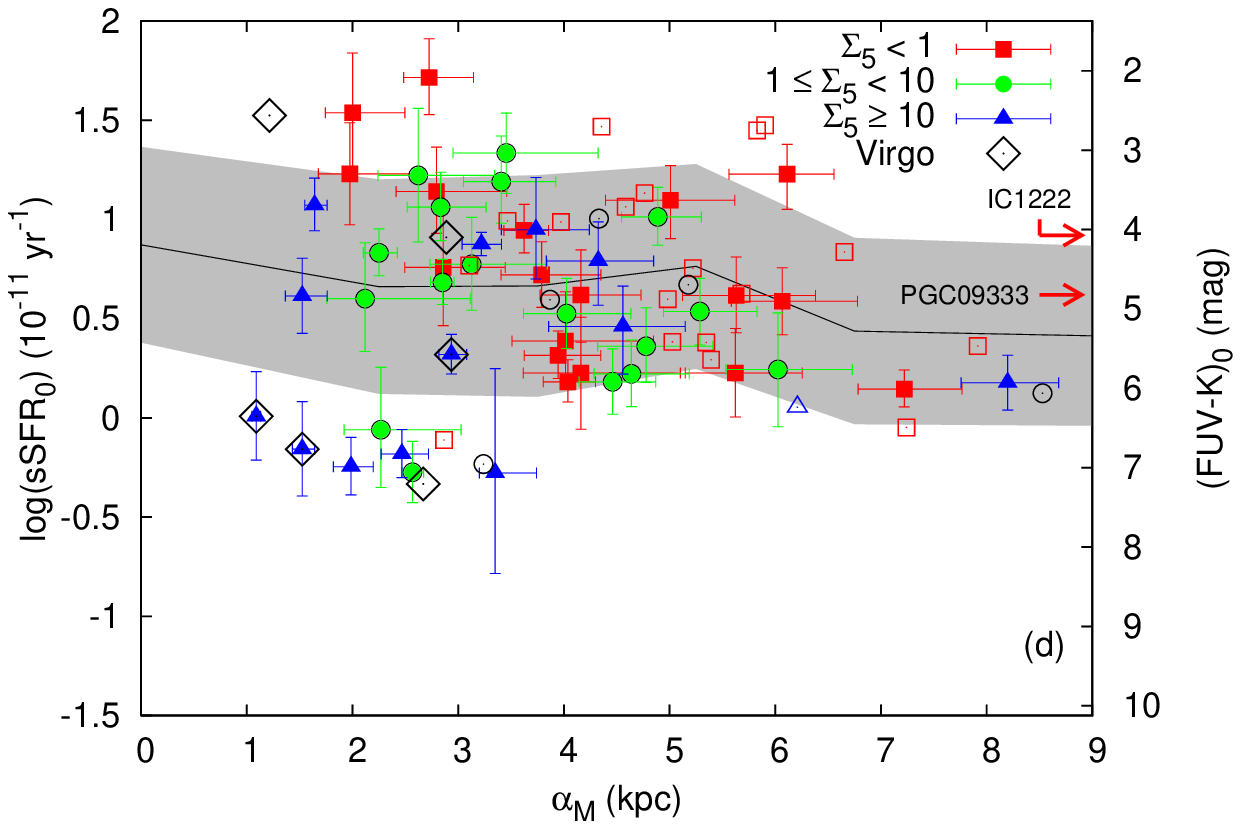}}\\
\resizebox{0.48\hsize}{!}{\includegraphics{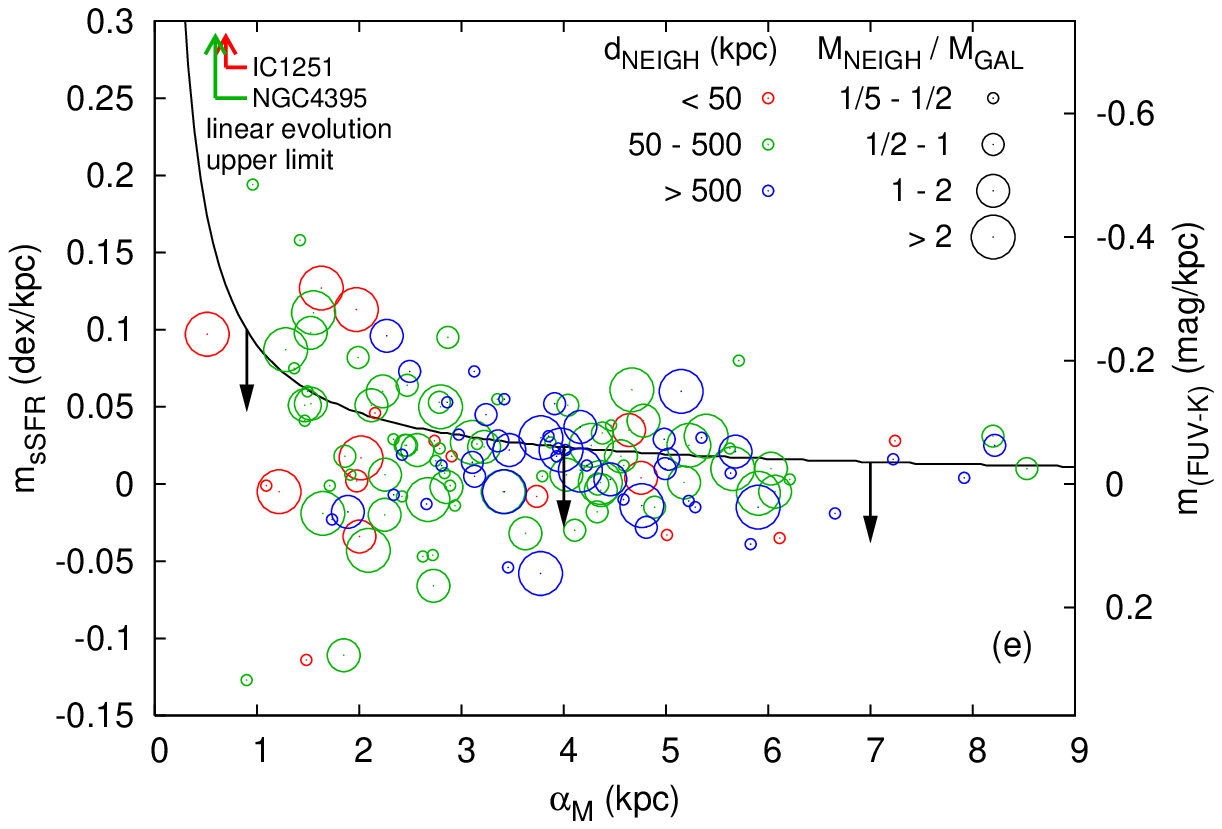}}\hspace{0.1cm}
\resizebox{0.48\hsize}{!}{\includegraphics{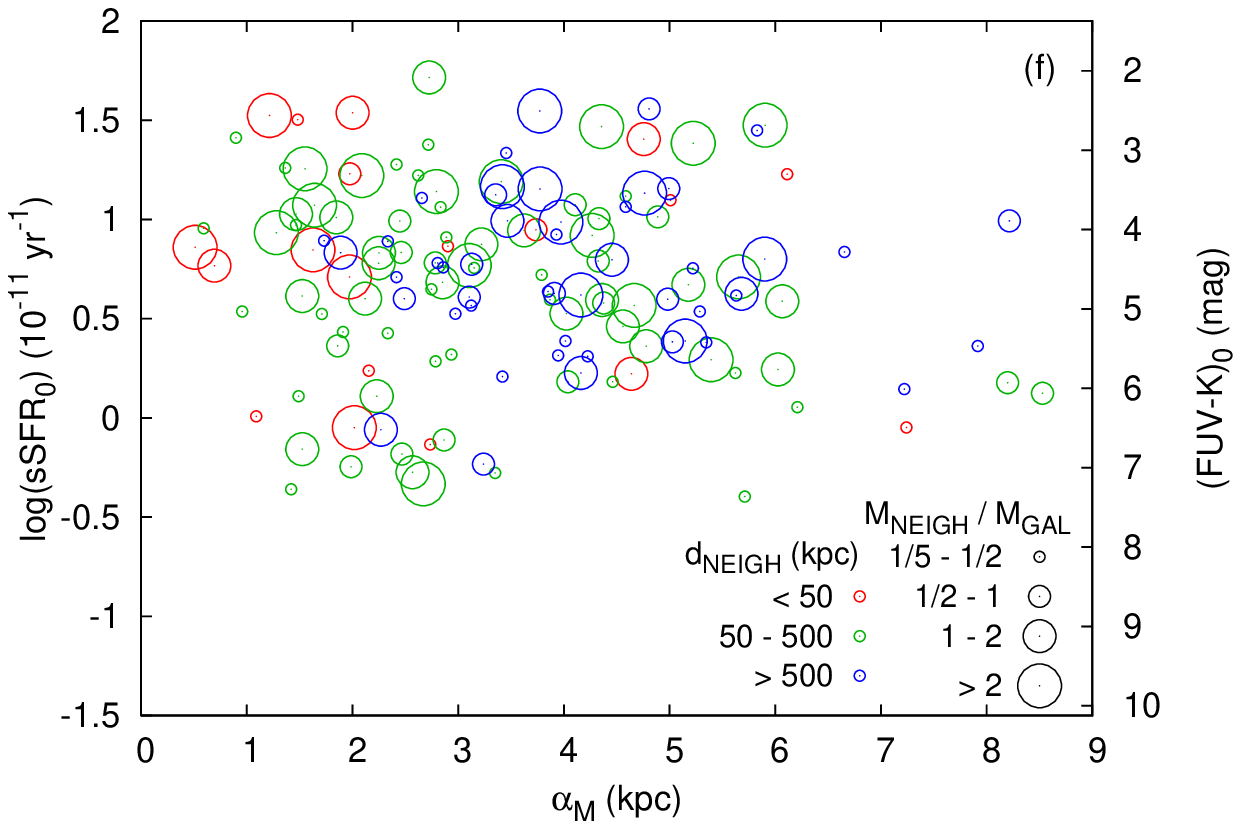}}\\
}

\clearpage
\begin{figure}
\figurenum{6}
\caption{(a) Specific SFR gradient [and the equivalent (FUV$-$$K$) color gradient] as a function of the scale-length of the stellar mass surface density profile. Different colors and symbols are used to sort out galaxies into three bins of absolute $K$-band magnitude (and mass, therefore). The black solid line and the gray shaded band show the mean value and standard deviation of $m_{sSFR}$ in bins of 1.5 kpc, computed from the whole set of randomly simulated values in each bin. Colored arrows mark the positions of galaxies out of range. Open symbols are used for galaxies with $\Delta\alpha_{M}>1$~kpc. (b) Same as (a), but with the extrapolated value of $\log(sSFR)$ at $r=0$ [and the corresponding (FUV$-$$K$) color]. (c,d): Same as (a) and (b), but with the galaxies segregated in three bins of projected local galaxy density, in units of Mpc$^{-2}$. Galaxies belonging to the Virgo cluster are marked with a black diamond. (e,f): Same as (a) and (b), but with galaxies sorted out according to the distance to and mass of the nearest neighbor with $M_{NEIGH} \geq 0.2 \times M_{GAL}$. The black line in panel (e) corresponds to the maximum sSFR gradient predicted by a linear evolution model with $\tau=\infty$ (see Sections~\ref{model} and \ref{discussion} as well as Fig.~7a).}
\end{figure}

\clearpage
\begin{figure}
 \figurenum{7}
\centering
\epsscale{1.0}
\resizebox{1\hsize}{!}{\includegraphics*[65,190][545,695]{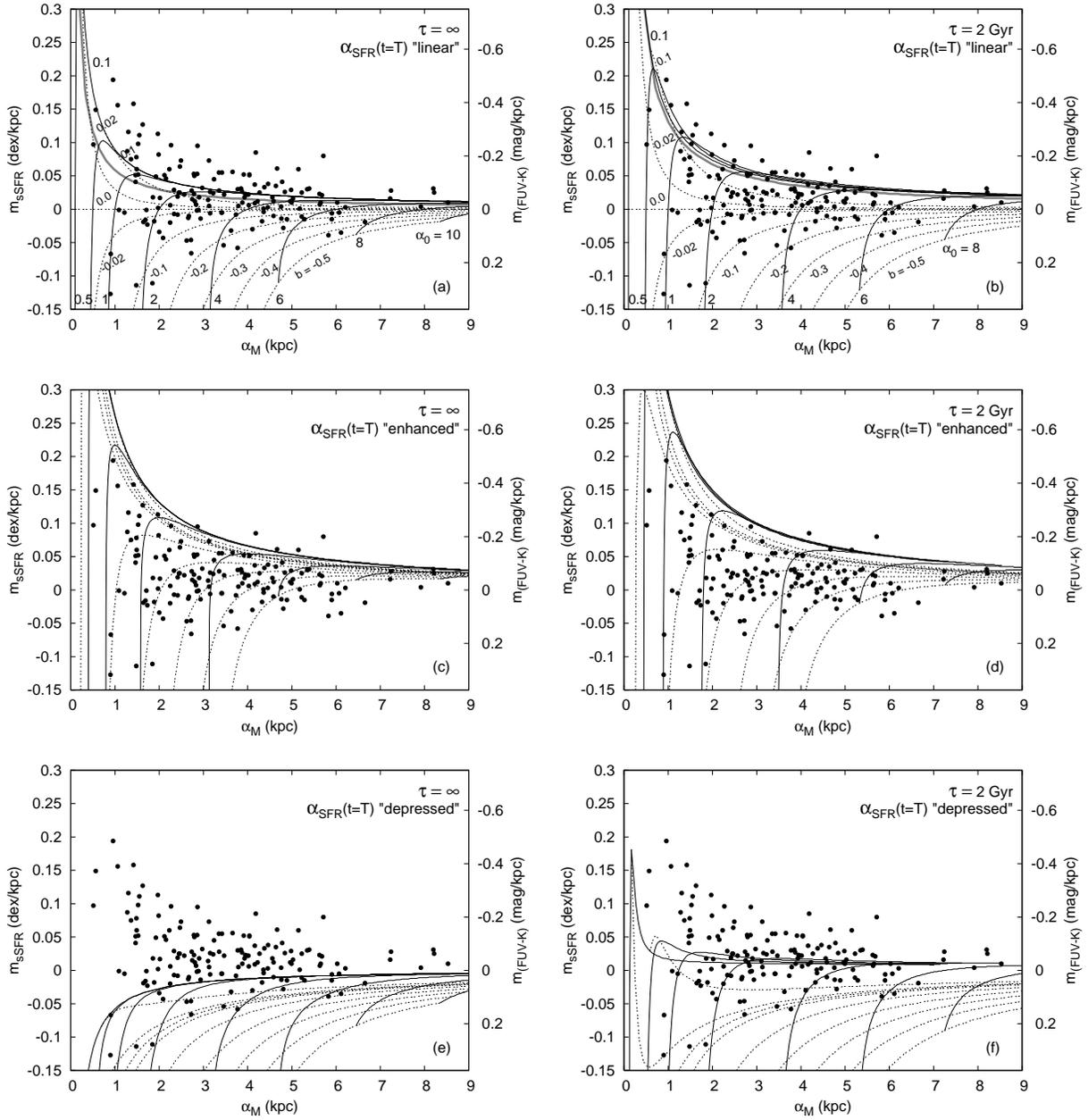}}
\caption{Model predictions of the sSFR radial gradient as a function of the scale-length of the stellar mass surface density profile. Black solid lines are curves with constant values of $\alpha_{0}$ (in kpc), and dashed lines have constant values of $b$ (in kpc/Gyr). The thick gray line corresponds to disks with $\alpha_{M}(T)=1.25\alpha_{0}$. (a) Model prediction for a $\tau=\infty$ SFR timescale and `linear' evolution. (b) Same as (a) but with $\tau=2$ Gyr. (c,d): Models with an `enhanced' recent outer-disk star formation (see text for details). (e,f): Models with a `depressed' recent outer-disk star formation (see text).}
\end{figure}

\end{document}